\documentclass[useAMS,usenatbib]{mn2e}
\usepackage{graphics}

\newcommand{\Ks}{\mbox{$K$$s$}}
\newcommand{\gmas}{\mbox{$\dot{M_g}$}} %gas mass loss rate
\newcommand{\mlu}{\mbox{$M_{\odot}$\,yr$^{-1}$}}
\newcommand{\msun}{\mbox{$M_{\odot}$}}
\def\oversim#1#2{\lower0.5pt\vbox{\baselineskip0pt \lineskip-0.5pt
     \ialign{$\mathsurround0pt #1\hfil##\hfil$\crcr#2\crcr\sim\crcr}}}
\def\gsim{\mathrel{\mathpalette\oversim>}}    % > over \sim
\begin{document}
\title[Gas and dust in the LMC]
{The global gas and dust budget of the Large Magellanic Cloud: AGB stars and supernovae, and the impact on the ISM evolution
%and comparisons with other gas and dust sources
}
%   \subtitle{Discrepancy of dust compositions between dust sources and ISM extinction curve}
\author[M. Matsuura et al.]
{M.~Matsuura$^{1,2}$, 
M.J. Barlow$^{2}$,
A.A.~Zijlstra$^{3}$,
P.A.~Whitelock$^{4,5,6}$,
M.-R.L. Cioni$^{7}$,
\newauthor
M.A.T. Groenewegen$^{8}$, 
K. Volk$^{9}$,
F.~Kemper$^{3}$,
T.~Kodama$^{1}$,
E.~Lagadec$^{3}$,
\newauthor
M.~Meixner$^{10}$,
G.C.~Sloan$^{11}$,
S. Srinivasan$^{12}$,  
%M.A.T.~Groenewegen$^{11}$?,
%P.R.~Wood$^{8}$?,
%A.G.G.M. Tielens?,
%K. Volk?,
%\newauthor
%E. Lagadec$^{3}$ ?,
%J.Th.~van~Loon$^{10}$ ?, 
%\newauthor
\\
%
%$^{2}$ APS Division, Department of Pure and Applied Physics, 
%        Queen's University Belfast, University Road, BT7 1NN, United Kingdom \\
$^{1}$ National Astronomical Observatory of Japan, Osawa 2-21-1, 
       Mitaka, Tokyo 181-8588, Japan \\
$^{2}$ Department of Physics and Astronomy, University College London, 
	Gower Street, London WC1E 6BT, United Kingdom \\
$^{3}$ Jodrell Bank Centre for Astrophysics, School of Physics and Astronomy, 
       University of Manchester, 
       Oxford Street, Manchester M13 9PL, \\ United Kingdom \\
$^{4}$ South African Astronomical Observatory, P.O.Box 9, 7935     
        Observatory, South Africa \\
$^{5}$  NASSP, Astronomy Department, University of Cape Town, 7701 Rondebosch, 
        South Africa \\
$^{6}$ Department of Mathematics and Applied Mathematics, 
        University of Cape Town, 7701 Rondebosch, South Africa \\        
$^{7}$  Centre for Astrophysics Research, University of Hertfordshire, Hatfield AL10 9AB, 
         United Kingdom \\
$^{8}$ Royal Observatory of Belgium, Ringlaan 3, B-1180 Brussels, Belgium \\
$^{9}$Gemini Observatory, Hilo, HI, USA \\
$^{10}$Space Telescope Science Institute, 3700 San Martin Drive, Baltimore, MD 21218, USA \\
$^{11}$ Astronomy Department, Cornell University, 610 Space Sciences Building, 
        Ithaca, NY 14853-6801, USA \\
$^{12}$Department of Physics and Astronomy, Johns Hopkins University, Homewood Campus, Baltimore, MD 21218, USA \\
%$^{8}$  Research School of Astronomy \& Astrophysics, Mount Stromlo Observatory,
%        Australian National University, 
%        Cotter Road, \\
%        Weston ACT 2611, Australia \\
%$^{11}$  Astrophysics Group, School of Physical and Geographical Sciences, Keele 
%        University, Staffordshire ST5 5BG, United Kingdom \\
%$^{10}$  Sterrewacht Leiden, Niels Bohrweg 2, 2333 RA Leiden, The Netherlands \\
%$^{11}$ Institut d'Astrophysique de Paris, CNRS, 98bis Boulevard Arago, 75014 Paris, France \\
        %$^{11}$ Universit\'{e} Paris 6, 98bis Bd Arago, 75014 Paris, France \\
%$^{13}$ Astronomical Institute ``Anton Pannekoek'', University of Amsterdam, 
%        Kruislaan 403, 1098 SJ, Amsterdam, \\
%        The Netherlands \\
             }

\date{Accepted 2009 March 5.  Received 2009 February 27; in original form 2008 October 29}
%\date{Accepted. Received; in original form }
\pagerange{\pageref{firstpage}--\pageref{lastpage}} \pubyear{2009}

\maketitle
\label{firstpage}
\begin{abstract}
  We report on an analysis of the gas and dust budget in the the interstellar
  medium (ISM) of the Large Magellanic Cloud (LMC).  Recent observations from
  the {\it Spitzer Space Telescope} enable us to study the mid-infrared dust
  excess of asymptotic giant branch (AGB) stars in the LMC.  This is the first
  time we can quantitatively assess the gas and dust input from AGB stars over
  a complete galaxy, fully based on observations.  The integrated mass-loss
  rate over all intermediate and high mass-loss rate carbon-rich AGB
  candidates in the LMC is $8.5\times10^{-3}$\,\mlu, up to
  $2.1\times10^{-2}$\,\mlu. This number could be increased up to
  $2.7\times10^{-2}$\,\mlu\, if oxygen-rich stars are included.  This is
  overall consistent with theoretical expectations, considering the star
  formation rate when these low- and intermediate-mass stars where formed, and
  the initial mass functions.

  AGB stars are one of the most important gas sources in the LMC, with
  supernovae (SNe), which produces about 2--4$\times10^{-2}$\,\mlu.  At the
  moment, the star formation rate exceeds the gas feedback from AGB stars and
  SNe in the LMC, and the current star formation depends on gas already
  present in the ISM.  This suggests that as the gas in the ISM is exhausted,
  the star formation rate will eventually decline in the LMC, unless gas
  is supplied externally.

  Our estimates suggest `a missing dust-mass problem' in the LMC, which is
  similarly found in high-z galaxies: the accumulated dust mass from AGB stars
  and possibly SNe over the dust life time (400--800\,Myrs) is significant less
  than the dust mass in the ISM.  Another dust source is required, possibly
  related to star-forming regions.
\end{abstract}
%
%________________________________________________________________
\large

%\tableofcontents

\section{Introduction}
The interstellar medium (ISM) of a galaxy is one of the drivers of its
evolution, and the composition of the ISM determines many of the
characteristics of the next generation of stars. The ISM is itself
continuously renewed and enriched by stellar ejecta. The enrichment occurs as
stars evolve and die, either exploding as supernovae (SNe) or experiencing
intense mass loss in a super-wind. Super-winds occur in low and intermediate
mass stars during the asymptotic giant branch (AGB) phase (main sequence
masses in the approximate range 1--8\,$M_{\odot}$), and in more massive red
supergiants. The ejecta are enriched with elements produced in various phases
of nuclear burning. In general terms, gas ejected from SNe includes newly
synthesised heavy elements, such as oxygen, iron and silicon
\citep{Nakamura99}, while AGB stars (below 8\,$M_{\odot}$) synthesize lighter
elements, especially carbon and nitrogen \citep{Maeder92}. The chemical
evolution of the gas can be well understood in terms of the different stellar
sources \citep[e.g.][]{CMR01}, taking into account the need for infall of
unprocessed gas to stabilize the final metallicity \citep{FD08}. The chemical
evolution of the Large Magellanic Cloud (LMC) has recently been studied by
\citet{Carrera08}.

The origin of the dust in the ISM is less well understood. Dust forms in
stellar ejecta at temperatures of about 1000--1500\,K and at high densities.
Different stars produce different types of dust. Red supergiants produce
oxygen-rich (silicate) dust.  AGB stars have two distinct chemical types:
oxygen- and carbon-rich, depending on the abundance ratio of oxygen and
carbon atoms within their atmospheres.  Oxygen-rich stars form silicate
dust, while carbon-rich stars yield amorphous carbon, graphite and SiC dust.
Supernovae can produce both dust types, depending on the abundances in the
different layers of the ejecta
\citep{Rho08}.

The current rate of ISM enrichment by dust and gas depends on the total
stellar population, the initial mass function and the star-forming history
\citep[e.g.][]{Salpeter55}.  Type II SNe are expected to dominate the
enrichment in the early phases of galaxy evolution \citep[e.g.][]{Maeder92}.
\citet{Hirashita02} argue that dust grains injected from SNe into the ISM
accelerate star formation in young galaxies. It takes more than 100\,Myr
for the first intermediate-mass stars to evolve onto the AGB 
\citep[e.g.][]{Vassiliadis93}.  Thus, dust and gas enrichment from
AGB stars occurs later than from high-mass stars.  Different galaxies,
at different stages of this process, may be expected to show differences in
gas-to-dust ratios, dust content, and, in consequence, ISM dust extinction
curves.

In our Galaxy, the major dust sources are presumed to be AGB stars and SNe
\citep{Gehrz89}. Some other sources, such as Wolf-Rayet stars and novae, also
contribute dust to the ISM of the Milky Way, but only in small quantities
\citep{Gehrz89}.  The relative importance of AGB stars and SNe remains
uncertain.  \citet{Dwek98} suggests that SNe are important silicate dust
sources, while most carbonaceous dust grains are from carbon-rich AGB stars. 
\citet{Jura89} show that AGB stars are an important gas source within our
Galaxy, except in the Galactic plane where SNe appear to dominate.
The dust formation efficiency in SNe remains controversial
\citep[e.g.][]{Sugerman06}, since SN shocks also destroy dust
\citep{Tielens94} and since dust production prior the explosion remains
unclear.

There are also uncertainties in the dust formation efficiencies for AGB stars.
This is expected to depend on the mass-loss rate, chemical type, and
metallicity. For oxygen-rich stars mass-loss rates decrease towards lower
metallicity \citep{Wood98, Bowen91, Marshall04}. In contrast, for carbon-rich
stars, although the metallicity dependence of mass-loss rates is still unclear
\citep{Habing96, vanLoon00, Wachter08, Mattsson08}, there does not appear to
be any obvious metallicity dependence in the range from solar, down to
one-twentieth of the solar metallicity \citep{Groenewegen07, Matsuura07, Sloan09}.
However, we would expect the mass-loss rates of carbon stars to differ between
environments with different abundances of alpha elements.  In these cases the
initial abundance of oxygen in the carbon-star precursor would be vital in
determining how much carbon must be dredged up before any was free to become
dust.

Galaxy evolution models aim at fitting the current composition of the ISM by
following the evolution of the stellar population over the lifetime of the
galaxy. This requires a detailed knowledge of dust formation efficiencies, gas
and dust expulsion efficiencies and nucleosynthesis in stellar interiors.
Here we aim to measure the current integrated rate of gas and dust input into
the ISM for one specific galaxy: the LMC.  Recent observations by the {\it
  Spitzer Space Telescope} (hereafter Spitzer \citep{Werner04}) facilitate the
detection of individual dust formation sites in nearby galaxies.  Observations
from Spitzer can thus be used to provide a new estimate of the gas and dust
budgets for these galaxies.

We have obtained spectra of AGB stars (mainly carbon-rich stars) in the LMC
and the Small Magellanic Cloud (SMC) using the Infrared Spectrograph
\citep[IRS; ][]{Houck04} on-board Spitzer \citep{Sloan06, Sloan08, Zijlstra06,
  Matsuura06, Wood07, Lagadec07, Leisenring08}; there are also independent
spectral surveys of these galaxies \citep{Buchanan06, Kastner08}.  A complete
LMC photometric survey using Spitzer was recently published by
\citet[][SAGE]{Meixner06}.  Combining these data we obtain a census of the
mass-losing stars, and can then compare AGB stars with other sources of dust
and gas in the LMC. The SMC data are not discussed in detail, but are used to
extend the colour mass-loss relations to lower mass-loss rates.

\section{Analysis}

\subsection{Mass-loss rates and colours}\label{colour-mass-loss}

%\include{lmc_table1}
%\include{smc_table2}

%_________________________________________________________________
\begin{figure}
\centering
\resizebox{\hsize}{!}{\includegraphics*{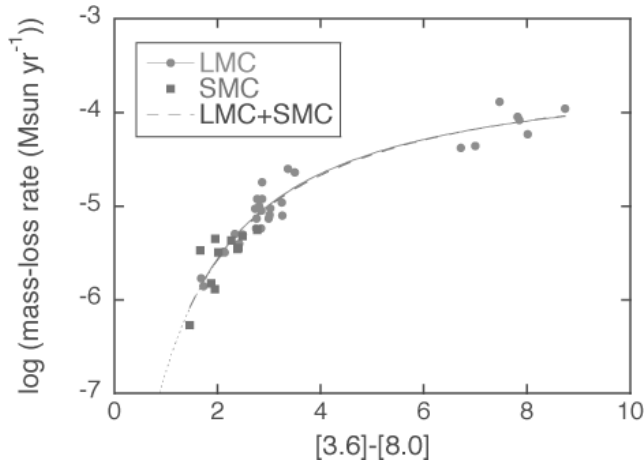}}
%\resizebox{\hsize}{!}{\includegraphics*[14,285][503, 649]{3_8_massloss2.eps}}
\caption{Gas mass-loss rates for LMC and SMC carbon stars from
  \citet{Groenewegen07} and \citet{Gruendl08} are plotted as a function of
  $[3.6]-[8.0]$ colours extracted from SAGE \citep{Meixner06}. A gas-to-dust
  ratio of 200 is adopted.  The solid (red) curve is the fit to the LMC sample
  only and the dashed (green) curve is the fit to the combined LMC and the SMC
  samples.  }
\label{Fig-3-8-massloss}
\end{figure}
%_________________________________________________________________
%_________________________________________________________________
\begin{figure}
\centering
\resizebox{\hsize}{!}{\includegraphics*{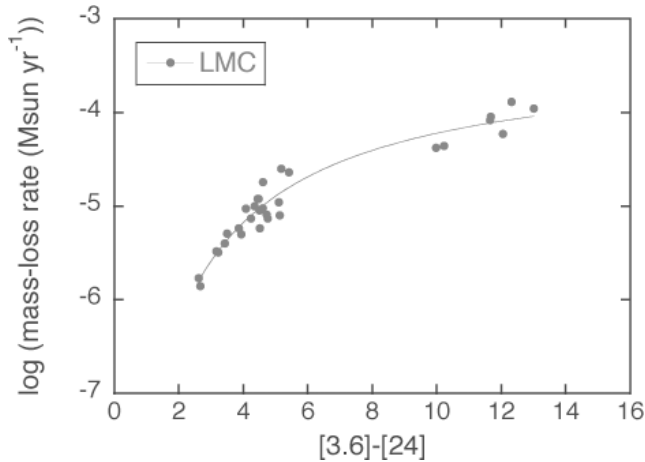}}
%\resizebox{\hsize}{!}{\includegraphics*[14,285][503, 649]{3_24_massloss2.eps}}
\caption{Similar to Fig.\,\ref{Fig-3-8-massloss}, showing $[3.6]-[24]$ for
  the LMC sample.}
\label{Fig-3-24-massloss}
\end{figure}
%_________________________________________________________________
%%_________________________________________________________________
%\begin{figure}
%\centering
%\resizebox{\hsize}{!}{\includegraphics*[14,285][503, 649]{k_24_massloss.eps}}
%\caption{Similar to Fig.\,\ref{Fig-3-8-massloss}, but for $\Ks-[24]$ for the
%  LMC sample.  }
%\label{Fig-k-24-massloss}
%\end{figure}
%%_________________________________________________________________
%_________________________________________________________________
\begin{figure}
\centering
\resizebox{\hsize}{!}{\includegraphics*{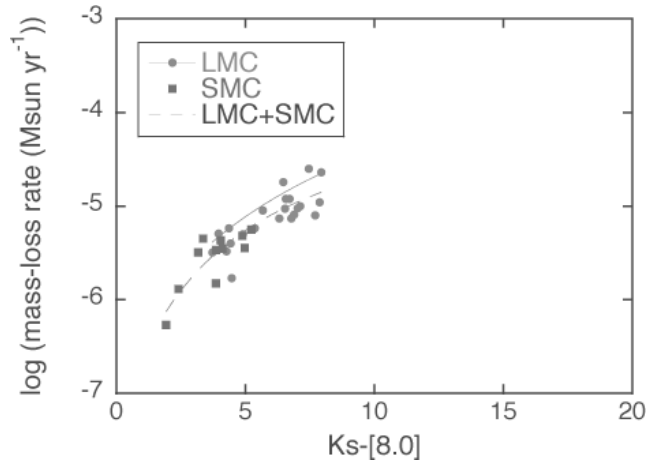}}
%\resizebox{\hsize}{!}{\includegraphics*[14,285][503, 649]{k_8_massloss2.eps}}
\caption{Similar  to Fig.\,\ref{Fig-3-8-massloss}, showing $\Ks-[8.0]$.
}
\label{Fig-k-8-massloss}
\end{figure}
%_________________________________________________________________

It has been shown empirically that mass-loss rates of AGB stars correlate
with their infrared colours \citep[e.g.][]{Jura87, Whitelock94, LeBertre97}. 
Particularly well known is the correlation with $K-[12]$ \citep{Whitelock06,
vanLoon99}, where [12] is the IRAS 12\,$\mu$m magnitude.  It is assumed that
this arises because $K$ magnitude is a measure of the flux from the central
star, while [12] measures the flux from the circumstellar envelope. A
similar mass-loss rate versus colour relation has been obtained using
[6.4]$-$[9.3] for carbon-rich AGB stars in the LMC and SMC
\citep{Groenewegen07}. \citet{Groenewegen06} also provide theoretical
relations between mass-loss rates and infrared colours for photometry from 
Spitzer, 2MASS \citep{Skrutskie06} and {\it AKARI}
\citep{Murakami07}.

We derive observational mass-loss rate versus colour relations by adopting gas
mass-loss rates estimated for carbon-rich AGB stars in the Magellanic Clouds
from \citet{Groenewegen07} and \citet{Gruendl08}.  \citet{Groenewegen07} used
the spectral energy distributions obtained from Spitzer IRS spectra
\citep{Sloan06, Zijlstra06} and $JHKL$ mags observed at Siding Spring
Observatory (SSO) in Australia, near simultaneously with the IRS observations
\citep{Wood07}.  The spectral energy distributions were fitted using a
radiative transfer model.  The gas-to-dust ratio is assumed to be 200 for
these carbon-rich AGB stars.  For all these stars, we extract Spitzer IRAC
\citep{Fazio04} and MIPS \citep{Rieke04} mags. For the LMC the Spitzer and the
2MASS mags are taken from the 2007 version of SAGE \citep[Surveying the Agents
of a Galaxy's Evolution; ][]{Meixner06}, while for the SMC they are from
S$^3$MC \citep{Bolatto07}.  Stars in clusters are excluded here so as to avoid
possible source confusion.  We cross-identified for sources within 10\,arcsec
from SAGE and S$^3$MC, using the coordinates quoted in \citet{Sloan06} and
\citet{Zijlstra06}.  We found 24 stars in the LMC and 19 in the SMC.  Some of
them have incomplete photometric data sets from 2MASS and/or SAGE. Highly
reddened stars were not detected in the 2MASS $J$-band.  \citet{Gruendl08}
discovered thirteen extremely red objects (EROs) in the LMC, and seven of them
were confirmed to be carbon-rich AGB stars, based on IRS spectra; we use their
mass-loss rates in the following analysis.  \citet{Gruendl08} measured their
magnitudes in the Spitzer IRAC bands and the MIPS 24 and 70\,$\mu$m bands, but these
stars were not detected in the near-infrared either by 2MASS
\citep{Skrutskie06} or by \citet{Kato07}.

The average interstellar extinction is estimated to be $E(B-V)$=0.15 and
$A_{Ks}$=0.04 mag towards the LMC and $A_{Ks}$=0.01 mag towards the SMC
\citep{Glass99, Zaritsky99, Cioni06a, Cioni06b}.  We can safely ignore the interstellar
extinction here, except for the $J-\Ks$ versus \Ks\,diagram in
Fig.\,\ref{Fig-JK-K}, for a comparison with a figure in \citet{Cioni06a}.

Figs.  \ref{Fig-3-8-massloss} -- \ref{Fig-k-8-massloss} show the mass-loss
rates as a function of [3.6]$-$[8.0], [3.6]$-$[24] and
\Ks$-$[8.0], respectively. These relations are fitted with a log function,
following \citet{LeBertre98}, and the fits are shown in
the figures.  Fits to the LMC and SMC combined sample give
\begin{equation} \label{eq-38}
\log\, \gmas =  -6.20 / ( ([3.6]-[8.0])+0.83) -3.39   
\end{equation}
in the range of $1<[3.6]-[8.0]<9$
\begin{equation} \label{eq-k8}
\log\, \gmas =  -14.50 / ( (Ks-[8.0])+3.86) -3.62  
\end{equation}
in the range of $1<Ks-[8.0]<9$,
and fits to the LMC sample ($3<[3.6]-[24]<14$) give
\begin{equation} \label{eq-324}
\log\, \gmas =  -9.89 / ( ([3.6]-[24])+1.43) -3.36 
\end{equation}
%\begin{equation}\label{eq-k24}
%log\, \gmas =  -20.25 / ( (Ks-[24])+4.98) -3.47,
%\end{equation}
where \gmas\, is the gas mass-loss rate.  These equations are fitted for stars
with colour ranges of $1.5 < [3.6]-[8.0]<9$ for eq.(\ref{eq-38})
$2<\Ks-[8.0]<8$\,mag for eq.(\ref{eq-k8}), $2.5<[3.6]-[24]<13$ for
eq.(\ref{eq-324}), respectively.  All fits have correlation coefficients
higher than 0.8. Coefficients are higher than 0.96 for $[3.6]-[8.0]$ and
$[3.6]-[24]$. The photometric data of the EROs contribute to the better fit in
the reddest colour range for the Spitzer bands, but these stars were not
detected in \Ks.  MIPS [24] data are used for the LMC only. Relatively low
mass-loss rate stars are not detected in MIPS [24] (Fig.\,\ref{Fig-824-8}),
and the equation using $[3.6]-[24]$ is suitable for relatively high mass-loss
rate stars only. Stars in the SMC sample have bluer infrared colours and lower
mass-loss rates than those of the LMC sample; this bias was present in the
original sample selection for IRS spectroscopy \citep{Sloan06, Zijlstra06,
  Lagadec07}.  If data in all bands are available the equation in
$[3.6]-[8.0]$ is the better one for deriving mass-loss rates, except for low
mass-loss rate stars, where $\Ks-[8.0]$ will be more sensitive to the small
dust excess. These two equations will be used later.

 These fits should be regarded as very uncertain in the extrapolated regime
of extremely low mass-loss rates.  Also, the equation fitted for $\Ks-[8.0]$ is
not appropriate for the mass-loss rates of red carbon stars
($>10^{-5}$\,$M_{\odot}$\,yr$^{-1}$).

Throughout this paper we assume a gas-to-dust ratio of 200 for carbon-rich
AGB stars to convert the measured dust mass to a gas mass-loss rate.
This ratio is similar to Galactic values \citep[c.f.][]{Habing96, LeBertre97,
  Skinner99}.  \citet{Matsuura05} argue that, for carbon-rich AGB stars, the
gas-to-dust ratio only marginally depends on the metallicity of the host
galaxy. This is because carbon atoms are synthesised in AGB stars, and the
quantity of amorphous carbon (or graphite) produced depends on the number of
excess carbon atoms in the atmosphere of the AGB star after all the oxygen
atoms are locked into the stable carbon monoxide molecule.  Various
observations suggest no obvious differences in mass-loss rates
for carbon-rich AGB stars amongst the LMC, the SMC and the Fornax dwarf
Spheroidal Galaxy \citep{Groenewegen07, Matsuura07}; the time variations and
uncertainties of the mass-loss rates appear to be larger than any metallicity
effects.  Typical metallicities of the LMC, the SMC, and the Fornax dwarf
Spheroidal Galaxy are half, one quarter and one tenth of the solar
metallicity, respectively.  Theoretical models by \citet{Mattsson08} and
\citet{Wachter08} predict that the mass-loss rates should depend on
metallicity, although the predicted difference between the LMC and SMC amounts
to only a factor of two.  \citet{Habing96} and \citet{Matsuura07} and
argued that the mass-loss rate of carbon stars is primarily
determined by the difference in the numbers of carbon and oxygen atoms: as the
excess carbon is self produced by the AGB star, carbon stars will show
relatively little difference in mass-loss rates between different galaxies,
although for low metallicity systems the mass loss may begin earlier on the
AGB phase \citep{LZ08}.
%The gas-to-dust ratio has been estimated for the Galactic carbon star,
%IRC+10\,216, \citep{LeBertre97} and should depend on the radius of the dust
%forming region, which remains uncertain.
Note that this will apply only to carbon-rich AGB stars;
the mass loss of oxygen-rich AGB stars is likely to show a strong
metallicity dependence \citep{Wood98, Bowen91, vanLoon00, LZ08}.

 In the process of estimating the mass-loss rate \citep{Groenewegen07,
    Gruendl08}, an outflow velocity of the circumstellar envelope is assumed.
  For oxygen-rich stars, lower outflow velocities are found \citep{Wood92,
    Marshall04}, but the gas velocity has never measured for carbon-rich stars
  in the LMC. Theoretical models \citep{Mattsson08, Wachter08} suggest that
  the mean velocity does not depend on metallicity for carbon-rich stars.
  \citet{Groenewegen07} assumed a gas velocity of 10\,km\,s$^{-1}$, while
  \citet{Gruendl08} used 8--12\,km\,s$^{-1}$.

 The other uncertainty originates from the assumption that the dust
  condensation temperature is 1000\,K \citep{Groenewegen06, Gruendl08}. This
  is a similar value as used for Galactic stars \citep{Groenewegen98}.
  However, for example, the condensation temperature of graphite dust grains
  varies with carbon-to-oxygen (C/O) ratio \citep{Lodders95, Tielens05}, and
  at a C/O ratio of 1.2, the temperature is 1700\,K.  It has been suggested
  that the C/O ratio of carbon-rich stars in the LMC is systematically higher
  than in their Galactic counterparts \citep{Matsuura05, Matsuura06}.  This
  has also been suggested from the observations of PNe in the LMC and the SMC
  \citep{Leisy96, Dopita97, Stanghellini05}.  To account for C/O ratio
  variations and to test their influence of the dust condensation temperature
  on mass-loss rate,
  we tested radiative transfer modelling of the SED fitting, with dust
  condensation temperatured of 1000\,K, 1200\,K, 1300\,K and 1700\,K.  We used
  the spectrum of one of the highest mass-loss rate stars,
  050231.49$-$680535.8 (J2000), observed by \citet{Gruendl08}.  The result
  shows that both the fitted optical depth and expansion velocity increase
  with dust condensation temperature.  The mass-loss rate required to fit the
  SED could be a factor of 2.4 higher, if the dust condensation temperature
  increases from 1000\,K to 1700\,K, although the numerical error is large for
  the models with dust condensation temperature above 1300\,K.  This suggests
  that the mass-loss rate, which is estimated from infrared colour, could be
  underestimated by a factor of up to 2.4, systematically.  In the overall
  analysis, we give the results under the assumption of 1000\,K dust
  condensation temperature, however, in the final gas and dust budget, we
  include this potential increase.

 AGB stars are long period variables, whose magnitudes vary over a
  100--1000 day time scale.  The amplitudes of the infrared variability are
  smaller than those of optical variabilities: they range from 0.2 to 1.5\,mag
  at $L$-band (3\,$\mu$m) \citep{LeBertre92, Whitelock06}.  Furthermore,
  mass-loss rates of AGB stars vary over time scales of 1000\,years
  \citep{Izumiura96, Leao06}.  The cause of long-term mass-loss rate
  variations are still unknown \citep{Zijlstra02}.  The variability affects
  our analysis for individual sourcs, but are averaged in the full, large
  sample.  

\subsection{Separation of oxygen-rich and carbon-rich AGB stars} \label{selection}

To estimate the integrated mass-loss rates from carbon-rich AGB stars, it is
essential to find a classification scheme that will separate carbon-rich from
oxygen-rich stars.  In particular, oxygen-rich AGB stars and red supergiants
(RSGs) follow different mass-loss rate vs. colour relations from carbon-rich
stars \citep{Whitelock06} and their dust contents will be very different.  

We first compiled a list of AGB stars in the LMC observed with the Spitzer IRS
by various groups \citep[][Kemper et al. in preparation]{Zijlstra06, Speck06,
  Kastner08, Leisenring08, Sloan08}.  These studies selected stars mainly from
previous LMC mid-infrared surveys, {\it IRAS} \citep{Schwering90}, {\it MSX}
\citep{Egan01}, and surveys for long-period variables \citep[e.g.][]{Hughes89,
  Hughes90}.  We used the carbon- and oxygen-rich classifications quoted in
these works. These are mostly based on the dust features (either silicate or
SiC at 10\,$\mu$m), but for stars without dust features, molecular bands are
used for chemical classifications.  The group of oxygen-rich stars includes
both giants and super-giants, but as shown in Sect\,\ref{other-colours}, two
young stellar objects (YSOs) or post-AGB stars with silicate bands may
contaminate the sample.  We searched the photometric data in SAGE to find the
counterparts to the stars observed with IRS: 122 stars have counterparts
within 1\,arcsec, 7 stars within 1--2 arcsec, and 5 stars have more distant
counterparts.  We use only identifications where the positions agree to within
2\,arcsec.  For the LMC this provides a total of 76 oxygen-rich and 40
carbon-rich stars from the SAGE database. We further add one oxygen-rich star
(IRAS 05298$-$6957) from an OH maser survey in the LMC \citep{Wood92}; the
other known OH maser sources were observed with IRS. In the following figures
the OH/IR star is included in `O-AGB/RGB (IRS)'.

A catalogue of spectroscopically identified optical carbon
stars in the LMC \citep{Kontizas01} was also searched for SAGE sources
% (Loup et al. 2003 document for positions of Kontizas 2001). 
within 3\,arcsec.  In practice, 98\, percent of the sources are found within
1\,arcsec, yielding 5710 SAGE counterparts.  Where there are more than two
SAGE sources within 3\,arcsec of the \citet{Kontizas01} coordinates, we
chose the closest one.

\citet{Cioni01} have obtained optical spectra of 126 red stars in the LMC,
and identified 11 carbon-rich and 61 oxygen-rich stars. These are described
as `C-AGB (optical)' and `O-AGB (optical)' in the figures. We also included
M-type stars (both giants and supergiants) from following works,
\citet{Sanduleak77}, \citet{Blanco80}, \citet{Westerlund81}, \citet{Wood83,
Wood85}, \citet{Hughes89}, \citet{Reid88} using
coordinates taken from  SIMBAD. There are 1710 objects, having 1568
counterparts in IRAC and 971 counterparts in MIPS within 1 arcsec. These
stars are identified as `O-AGB/RGB (Sim)' in the figures, although some of
them could be red giant branch (RGB), rather than AGB stars.

 \subsubsection{Classifications involving the 8\,$\mu$m band}

%_________________________________________________________________
\begin{figure*}
\centering
\rotatebox{90}{ 
%\begin{minipage} {6.5cm}
\begin{minipage} {11cm}
%\resizebox{\hsize}{!}{\includegraphics*[11,20][523,679]{mag38_8_nomark_lo.ps}}
%\resizebox{\hsize}{!}{\includegraphics*[11,20][523,679]{mag38_8_nomark.ps}}
\resizebox{\hsize}{!}{\includegraphics*[64,120][502,686]{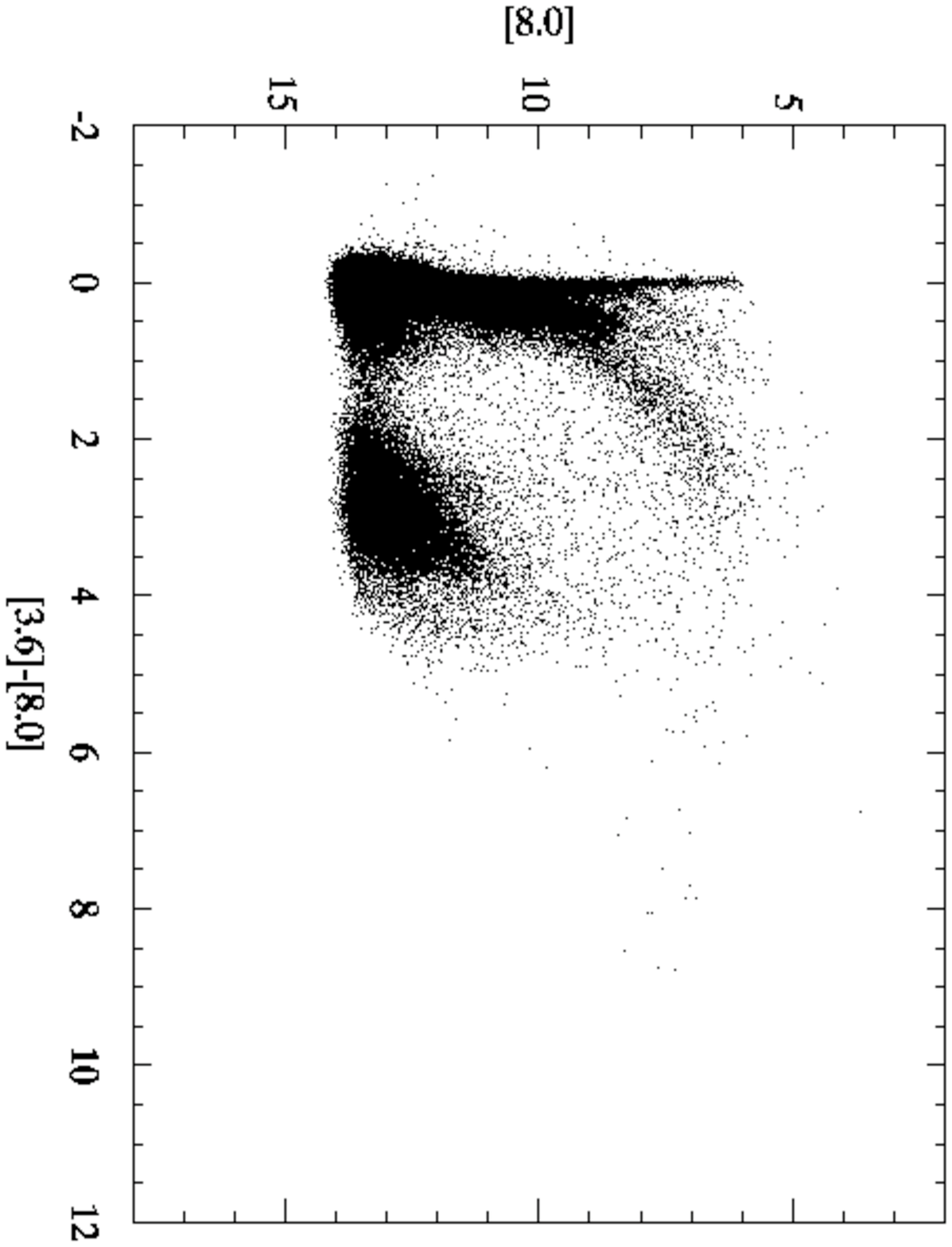}}
\end{minipage}}
\rotatebox{90}{ 
%\begin{minipage} {6.5cm}
\begin{minipage} {11cm}
%\resizebox{\hsize}{!}{\includegraphics*[11,20][523,679]{mag38_8_lo.ps}}
\resizebox{\hsize}{!}{\includegraphics*[11,20][523,679]{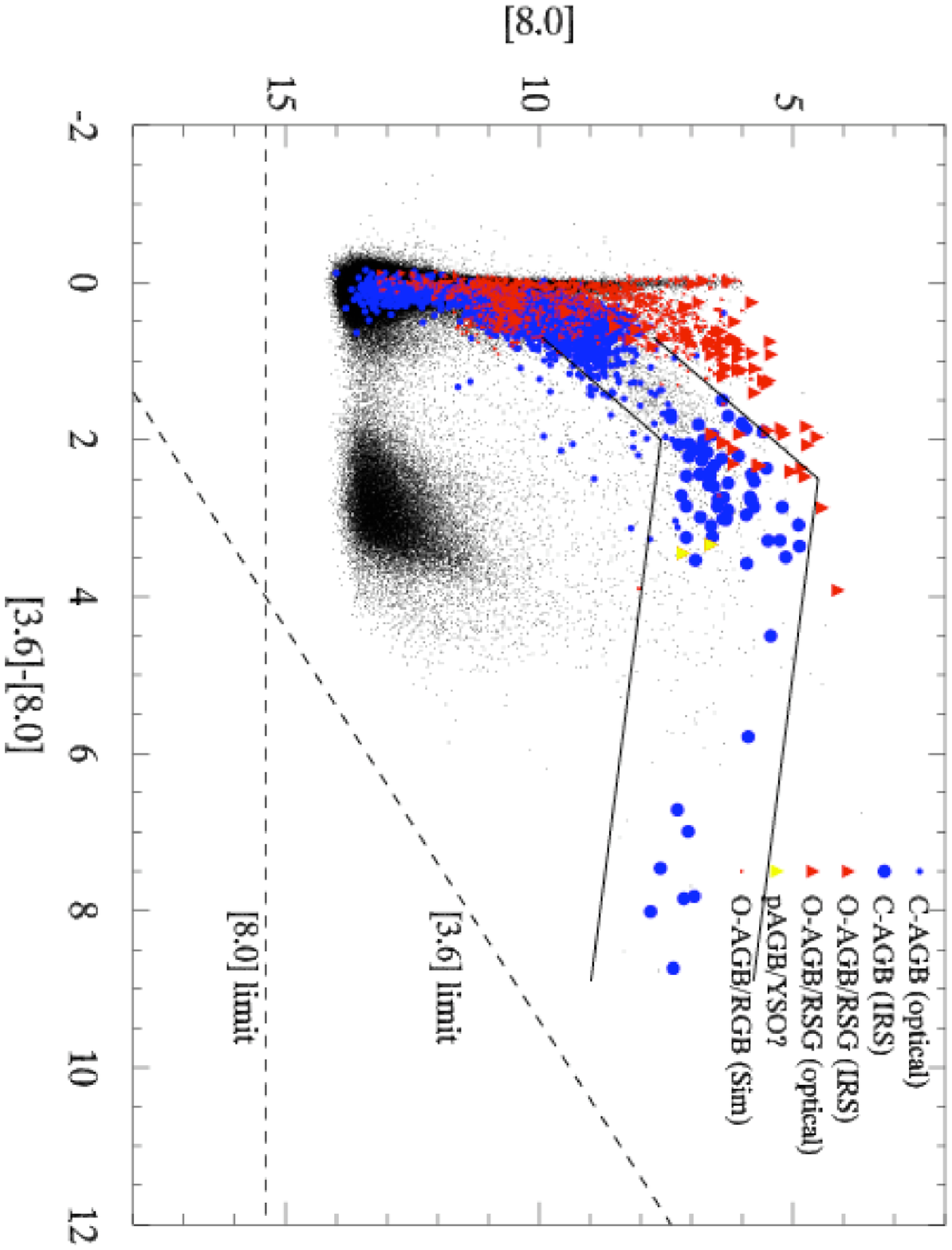}}
%\resizebox{\hsize}{!}{\includegraphics*[11,20][523,679]{mag38_8_lo.ps}}
\end{minipage}}
\caption{
$[3.6]-[8.0]$ vs [8.0] colour-magnitude diagram.
The top panel shows the colour magnitude diagram for the complete SAGE
sample. In the lower panel, spectroscopically identified oxygen-rich and
carbon-rich AGB stars are plotted. The selection criteria used to identify
carbon-rich AGB candidates are indicated by the solid lines. The sensitivity
limits quoted by \citet{Meixner06} for the final data product are indicated
by the dashed lines. 
$[8.0]\sim10.5$ likely corresponds to the tip of the RGB branch
the number of carbon stars decreases at about
[3.6]$-$[8.0]$\sim$0.7 (at about [8.0]$\sim$9), As seen in the top panel,
\label{Fig-38-8}}
\end{figure*}
%_________________________________________________________________

In this section we develop a classification scheme using the $[3.6]-[8.0]$
colour to separate carbon-rich from oxygen-rich stars. Diagrams involving
other colours can be found in the Appendix. Fig.\,\ref{Fig-38-8} shows a
$[3.6]-[8.0]$ vs [8.0] colour magnitude diagram. The [8.0]-magnitudes of
carbon-rich stars peak at $[3.6]-[8.0]=$2--3\,mag. The
SED of thick circumstellar shells peaks longward of about 7\,$\mu$m. 

In Fig.\ref{Fig-38-8} oxygen-rich stars tend to show brighter
[8.0]-mags than carbon-rich stars at any given $[3.6]-[8.0]$, with the
exception of the two oxygen-rich post-AGB/YSO candidates.
\citet{Meixner06}; \citet{Blum06} used the $[3.6]-[8.0]$ vs [8.0] diagram
for source classification. We also classify `extreme-AGB
stars' \citep{Blum06} as oxygen- or carbon-rich, based on our
Spitzer IRS spectra.  The boundaries which we use to separate carbon-
from oxygen-rich AGB stars in Fig.\ref{Fig-38-8} are:
\begin{equation} \label{eq3}
[8.0]> 9.0 - 1.8\times [3.6]~~\rm{for}~0.7<[3.6]<2.5 
\end{equation}
\begin{equation}
[8.0]< 11.2 - 1.8\times [3.6]~~\rm{for}~0.7<[3.6]<2.0
\end{equation}
\begin{equation}
[8.0]> 4.0 - 0.2\times [3.6]~~\rm{for}~2.0>[3.6]
\end{equation}
and
\begin{equation}  \label{eq6}
[8.0]< 7.2 - 0.2\times [3.6]~~\rm{for}~2.5>[3.6].
\end{equation}
This region is in general consistent with the 
evolutionary tracks of carbon-rich stars
\citep{Marigo08}.

In addition to low luminosity oxygen-rich stars
(Sect.\,\ref{other-colours}), nine IRS classified oxygen-rich stars are
found in the carbon-rich region. Four of them are known to be OH/IR stars
\citep{Wood92} and their IRS spectra show the silicate band in absorption.
It is not possible to make a perfect division of the sources by this method,
and inevitably some oxygen-rich sources are found in the carbon-rich region. 
However, this colour selection provides a method of isolating the majority
of carbon-rich stars.  We note that the high mass-loss rate OH/IR stars
found in the carbon rich region are not expected to occur in large numbers.
Furthermore, carbon-rich stars with colours similar to the OH/IR stars have
lower mass-loss rates than the OH/IR stars, due to their different dust
properties.  Thus, the influence of blended oxygen-rich stars would be only
a small perturbation on our estimate of the integrated mass-loss rate from
carbon stars.

There are a large number of so called `optical' carbon-rich stars outside
the region defined by equations (\ref{eq3}--\ref{eq6}).  However, with
the available colour-colour or colour-magnitude diagrams we are unable to
separate oxygen- and carbon-rich stars with thin circumstellar
envelopes.  Mass loss from carbon-rich stars with thin shells is
discussed separately (Sect.\,\ref{mass-loss}).

At around $[3.6]-[8.0]>4$, a small number of unresolved galaxies might intrude
into the AGB star region.  \citet{Blum06} investigated the $[3.6]-[8.0]$
colours of distant galaxies.  They found galaxy candidates in the red and low
luminosity regime, concentrating at $[3.6]-[8.0]\sim3$ and $[8.0]>10$ (their
fig.~7).  They suggested that galaxy candidates extend up to $[8.0]\sim7$ at
$[3.6]-[8.0]\sim3$.  They did not classify point sources with
$[3.6]-[8.0]\sim3$ and $[8.0]<10$. However, our sample based on Spitzer IRS
spectra clearly shows the presence of carbon-rich stars at $[8.0]\sim7$ at
$[3.6]-[8.0]\sim3$.  We argue in Sec.\ref{distribution} that the majority of
objects with $[8.0]<7$ and $[3.6]-[8.0]>3$ belong to the LMC. From this we
might also deduce that the most populous bright near- and mid-infrared sources
in galaxies are AGB stars.  This will impact the infrared colour of galaxies
\citep{Tonini08}.

There is a change in the number density of carbon stars at Ks$-$[8.0]$\sim$1.4
(top panel of Fig.\,\ref{Fig-K8-K}) and $[3.6]-[8.0]\sim0.7$ (top panel of
Fig.\,\ref{Fig-38-8}). This blue/red colour separation might correspond to AGB
stars with and without significant mass-loss rates. For the bluer sources the
opacity includes molecular bands and H$^{-}$ continuous absorption.  For the
redder sources, dust emission contributes to the mid-infrared excess.  Indeed,
IRAS observations show a $[12]-[25]$ gap between $-1.2$ and $-1.4$
\citep{vanderVeen88} and stars redder than $-1.4$ are those with dust-driven
mass loss\footnote{ In the van der Veen et al. study the IRAS bands are not
  normalized with zero-magnitudes, but $[12]-[25]$ is defined by $2.5 \log
  (F_{25}/F_{12})$, where $F_{12}$ and $F_{25}$ are the IRAS fluxes in Jy in
  the 12 and 25\,$\mu$m bands.}. The gap can be understood as the separation
of huge numbers of stars with very little mass loss from those with
circumstellar envelopes. Once dust grains have condensed, mass-loss rates
increase drastically \citep{vanderVeen88}, as do colours.  The $\Ks-[8.0]$ and
$[3.6]-[8.0]$ do not show the obvious gap seen in $[12]-[25]$, as the
influence of dust on the [8.0]-band is not strong, but it is likely that the
decrease in stellar density is related to the increase in mass-loss rate.

Possible sources of contamination for our carbon-rich AGB sample include
H{\small II} regions.  \citet{Kastner08} have classified bright objects in the
MSX A-band, which covers 8\,$\mu$m, and they found four H{\small II} regions
in the range $4.0<A<6.5$ and $6<K-A<9$. Although the MSX $A$- and Spitzer
[8.0]-bands are not identical, the wavelength coverage is similar, suggesting
H{\small II} regions could be a contaminant.  A search for
cross-identifications of SIMBAD sources within three arcsec revealed one
H{\small II} region, five Be stars or emission-line objects, two YSOs, three
molecular clouds, one symbiotic star (essentially an AGB star with a compact
companion), two planetary nebulae (PNe), and one WR star.  \citet{Whitney08}
showed that luminous YSOs cross the AGB sequence, but we expect the high mass
YSO population to be significantly smaller than that of AGB stars. The initial
masses of carbon-rich AGB stars are about 1.5--5.0\,$M_{\odot}$
\citep{Vassiliadis93}, while YSOs, which can reach $[8.0]<8$, have masses
higher than 10\,$M_{\odot}$.  Using the IMF from \citet{Kroupa01}, with an
exponent of $\alpha=1.3$, and assuming a constant star formation rate, stars
with 1.5--5.0\,$M_{\odot}$ have a birth rate ten times higher than stars with
masses greater than 10\,$M_{\odot}$, although the SFR might vary
\citep{Feast95}. The lifetime of such high mass YSO is also much shorter than
that of the AGB phase. Similarly WR stars have high initial-masses and their
numbers should be low. PNe (and most symbiotic stars) are usually located
outside of our carbon-rich region, and are more concentrated at lower
[8.0]-band luminosities \citep[][who used the most complete PN catalogue from
\citet{Reid06}]{Hora08}.  We expect that because their lifetimes are short,
the PN contamination is low.
 
The sensitivity limits of the surveys are shown on the colour-magnitude
diagrams.  These sensitivities (5\,$\sigma$ detection limits) are those
for the final product taken from \citet{Skrutskie06} and
\citet{Meixner06}; note that the actual detection limits of the catalogues
used could be lower than those illustrated here.  The Spitzer
SAGE project includes observations of the LMC at two epochs, but the 
catalogue used (2007 Winter version) contains only data from the first of epoch
\citep{Meixner06}.  In Fig.\,\ref{Fig-JK-K}, the sensitivity (5\,$\sigma$
detection) limits of the 2MASS survey \citep{Skrutskie06} are plotted on
the diagram. The 2MASS sensitivity is not high enough to detect the reddest
part of the carbon-rich sequence.

\section{Results}

%_________________________________________________________________
\begin{table*}
% \centering
\caption{LMC carbon-rich AGB candidates with high mass-loss rate ($>1\times10^{-5}$\,\mlu). 
The full table is available in on-line.
  \label{highmassloss}}
%\begin{center}
 \begin{tabular}{ccccrrcccccc}
  \hline
 SAGE name              &    J      &   H      &   K      &   [3.6]  &   [4.5]  &    [5.8] &    [8.0] &    [24]  &    \gmas$_2$ \\
                     &  (mag)    & (mag)    & (mag)    &   (mag)  & (mag)    & (mag)    & (mag)    &    (mag) &   (\mlu) \\ \hline
 J045013.14$-$693356.7 &    15.681 &   15.784 &   14.348 &   12.182 &   10.938 &    9.537 &    7.764 &    3.274 &    $2.7\times10^{-5}$   \\
 J045358.59$-$691106.3 &    ---    &   15.774 &   14.194 &   11.392 &   10.603 &    8.865 &    7.273 &    1.008 &    $2.3\times10^{-5}$   \\
 J045842.03$-$680715.3 &    ---    &   ---    &   ---    &   10.323 &    8.539 &    7.078 &    5.765 &    3.695 &    $2.9\times10^{-5}$   \\
 J045935.79$-$682444.7 &    15.205 &   14.491 &   13.567 &   10.398 &    9.328 &    8.307 &    7.278 &    6.319 &    $1.1\times10^{-5}$   \\
 J043257.38$-$692633.0 &    ---    &   ---    &   14.323 &   10.124 &    8.688 &    7.505 &    6.453 &    4.535 &    $1.7\times10^{-5}$   \\
 J045613.48$-$693230.9 &    ---    &   ---    &   ---    &   10.637 &    8.915 &    7.557 &    6.363 &    4.143 &    $2.5\times10^{-5}$   \\
 J045845.98$-$682037.7 &    ---    &   ---    &   ---    &   10.741 &    8.877 &    7.403 &    6.076 &    3.662 &    $3.1\times10^{-5}$   \\
 J044854.41$-$690948.1 &    ---    &   ---    &   13.880 &   11.530 &   10.646 &    9.245 &    7.594 &    1.656 &    $2.1\times10^{-5}$   \\
 J044650.41$-$675124.2 &    ---    &   ---    &   13.956 &   10.428 &    9.227 &    8.274 &    7.373 &    5.796 &    $1.0\times10^{-5}$   \\
 J045433.84$-$692036.1 &    14.357 &   13.594 &   12.414 &    9.947 &    8.958 &    7.982 &    6.606 &    3.904 &    $1.3\times10^{-5}$   \\
\hline
\end{tabular}
\end{table*}
%________________________________________________________________

%_________________________________________________________________
\begin{table}
% \centering
  \caption{ Number distribution of mass-losing carbon-rich candidates in 
the LMC.  Two different colours ($\Ks-[8.0]$ and $[3.6]-[8.0]$)  were used
to estimate the mass-loss rates of individual stars. In the last column,
one or other of these colours were used, i.e., if $[3.6]-[8.0]<2.2$, $\Ks-[8.0]$ vs
mass-loss rate relation was adopted, otherwise the $[3.6]-[8.0]$ vs
mass-loss rate relation was used. The total gas mass-loss rate is obtained
by integrating the mass-loss rates of individual stars.
  \label{table-lmc-massloss}}
\begin{center}
 \begin{tabular}{cccccccc}
  \hline
\gmas\,range &  \multicolumn{2}{c}{\gmas$_1$} & \multicolumn{2}{c}{\gmas$_2$}\\
& $N$ & Total \gmas & $N$ & Total \gmas\\
{\tiny (\mlu)} & \multicolumn{2}{r}{\tiny ($10^{-4}$\mlu)} & \multicolumn{2}{r}{\tiny ($10^{-4}$\mlu)} \\\hline
~~~~~~~~~~\gmas$<$1$\times$$10^{-6}$ & 996  &  ~3$^{\ddag}$  & 588  & ~4$^{\dagger}$ \\
1$\times$$10^{-6}$$<$\gmas$<$3$\times$$10^{-6}$ & 298  &  ~5  & 522  & 10 \\
3$\times$$10^{-6}$$<$\gmas$<$6$\times$$10^{-6}$ & 188  &  ~8  & 352  & 14 \\
6$\times$$10^{-6}$$<$\gmas$<$1$\times$$10^{-5}$ & 109  &  ~9  &122   & ~9 \\
1$\times$$10^{-5}$$<$\gmas$<$3$\times$$10^{-5}$ & 140  & 24   &140  & 24 \\ 
3$\times$$10^{-5}$$<$\gmas$<$6$\times$$10^{-5}$ & ~39  & 15   & ~39    & 15 \\ 
6$\times$$10^{-5}$$<$\gmas$<$1$\times$$10^{-4}$ & ~13  & 10   & ~13    & 10 \\ \hline
Total &  1779 & 74$^\S$ & 1779 & 86$^\S$ \\
\hline
\end{tabular}
\end{center}
\gmas : gas mass-loss rates, assuming a gas-to-dust mass ratio of 200 \\
$N$: number of stars \\
\gmas$_1$ : \gmas\, estimated from $[3.6]-[8.0]$ only \\
\gmas$_2$ : \gmas\, estimated from $\Ks-[8.0]$ and $[3.6]-[8.0]$  \\
$^{\ddag}$ The minimum mass-loss rate is $4\times10^{-8}  M_{\odot}$ yr$^{-1}$, limited by $[3.6]-[8.0]>0.7$\,mag \\
$^{\dagger}$ The minimum mass-loss rate is $4\times10^{-8}  M_{\odot}$ yr$^{-1}$, limited by $\Ks-[8.0]>1.4$\,mag (as well as $[3.6]-[8.0]>0.7$\,mag). \\
$^\S$ Due to rounding,  adding \gmas \, in seven ranges does not give the identical number as the total  \gmas. 
\end{table}
%________________________________________________________________

%_________________________________________________________________
\begin{table}
% \centering
  \caption{Number distribution of LMC carbon-rich candidates as a 
function of mass-loss rates based on  $[3.6]-[8.0]$ and $\Ks-[8.0]$ colours in the entire SAGE survey area,
   the bar region, the bar centre, and the remainder of the surveyed area except of the bar region (quoted as `disk')
   \label{table-sp-dist-lmc}}
\begin{center}
 \begin{tabular}{lrrrrrrrrccccccccc}
  \hline
 \gmas$_2$ \,range & \multicolumn{4}{l}{Number fraction of stars} \\
 & All & Bar & Bar cent. & Disk \\
(\mlu)&  &  &   & \\ \hline
4$\times$$10^{-8}$$<$\gmas$<$1$\times$$10^{-6}$ & 33\,\%  & 32\,\% & 35\,\% & 35\,\%\\
1$\times$$10^{-6}$$<$\gmas$<$3$\times$$10^{-6}$ & 31\,\%  & 31\,\% & 32\,\% & 31\,\%\\
3$\times$$10^{-6}$$<$\gmas$<$6$\times$$10^{-6}$ & 18\,\%  & 19\,\% & 19\,\% & 17\,\%\\
6$\times$$10^{-6}$$<$\gmas$<$1$\times$$10^{-5}$ & 7\,\%    & 7\,\%   & 6\,\%     &7\,\%\\
1$\times$$10^{-5}$$<$\gmas$<$3$\times$$10^{-5}$ & 8\,\%    & 9\,\%   & 7\,\%     & 7\,\%\\
3$\times$$10^{-5}$$<$\gmas$<$6$\times$$10^{-5}$ & 2\,\%    &  2\,\%  & 2\,\%     &  2\,\%\\ 
6$\times$$10^{-5}$$<$\gmas$<$1$\times$$10^{-4}$ & 0.7\,\%    &  0.8\,\%  & 0.8\,\%     &   0.7\,\%\\ \hline
Total $N$ & 1779 & 1194 & 517 & 585 \\ \hline
\end{tabular}
\end{center}
 \gmas$_2$ : gas mass-loss rates estimated using  $\Ks-[8.0]$ and $[3.6]-[8.0]$ \\
 $N$: number of stars \\
 Percentages over the seven ranges do not sum to 100\,\% due to rounding.
\end{table}
%________________________________________________________________

%_________________________________________________________________
\begin{table*}
% \centering
  \caption{ Overall properties of the LMC  \label{table-sp-dist-lmc-prop}}
\begin{center}
 \begin{tabular}{lrrrrrrrrccccccccc}
  \hline
 & All surveyed area & Bar & Bar cent & Disk\\
 \hline
Total number of carbon-rich AGB candidates & 1779~~ & 1194~~ & 517~~ & 585~~ \\ 
Integrated mass-loss rate$^\ddag$ ($\times10^{-4}$\,\mlu) & 86~~ & 59~~ & 23~~ &   27~~ \\
Area (deg$^2$)   & 50.4 & 18.7 & 5.7 & 31.7 \\
Mass-loss rate per unit area ($\times10^{-4}$\,\mlu\,deg$^{-2}$) & 1.7 & 3.2 & 4.1 &  0.8\\
Same as above but in different units ($\times10^{-4}$\,\mlu\,kpc$^{-2}$) & 2.2 & 4.0 & 5.1 & 1.1  \\
\hline
\end{tabular}
\end{center}
The mass-loss rates from carbon-rich AGB stars could be underestimated by a 
factor of  up to 2.4, due to the unknown dust condensation temperature 
(Sect\,\ref{colour-mass-loss}).$^\ddag$  gas mass-loss rates are 
estimated using  $\Ks-[8.0]$ and $[3.6]-[8.0]$.
\end{table*}
%________________________________________________________________

\subsection{Mass-loss rates for all carbon-rich AGB candidates} \label{mass-loss}

The mass-loss rates for carbon-rich stars are estimated using the colour vs
mass-loss-rate relations derived in Sect.\,\ref{colour-mass-loss}.  The 
stars with high mass-loss rates are listed in Table\,\ref{highmassloss}. We
assume that statistically these relations will give an integrated mass-loss
rate for all carbon-rich mass-losing AGB candidates.  Specifically, we use
$[3.6]-[8.0]$, supplemented by
$\Ks-[8.0]$ for low mass-loss rate stars, to estimate the rates.

There are  1779 carbon-rich AGB candidates with $[3.6]-[8.0]>0.7$\,mag.  The
integrated mass-loss rate from all 1779 stars, derived from this colour, is
$7.4\times10^{-3}$\,\mlu.  The highest individual mass-loss rate is
$9.3\times10^{-5}$\,\mlu.  The upper limit may be set by the detection limit
in the [3.8]-band (Figs.\,\ref{Fig-38-8} and \ref{Fig-3-24}).
The distribution of stars amongst the mass-loss rate ranges is summarized in
Table\,\ref{table-lmc-massloss}.

For stars with low mass-loss rates, the fits to colour vs mass-loss rate using
the two different colours, $\Ks-[8.0]$ and $[3.6]-[8.0]>0.7$, give slightly
different results; this is because towards blue colours, the extrapolated
relations are poorly defined.  $\Ks-[8.0]$ is probably better for these low
mass-loss rate stars.  We combine the two estimates of the mass-loss rate as
follows: We adopt the estimate from $[3.6]-[8.0]$ for stars with
$[3.6]-[8.0]>2.2$, otherwise we take the rates from $\Ks-[8.0]$. If a star
does not have a \Ks-band magnitude, we use the rates from $[3.6]-[8.0]$.
Above $[3.6]-[8.0]>2.2$ ($\sim3\times10^{-6}$\,\mlu) up to \Ks\, detection
limit, the mass-loss rates from the two colours are essentially identical.
This procedure gives an integrated mass-loss rate of $8.6\times10^{-3}$\,\mlu,
as summarized in the last column of Table\,\ref{table-lmc-massloss}. This
combined method only makes a 10\,\% difference in the integrated mass-loss
rate from that using the $[3.6]-[8.0]$ relation alone, as the low mass-loss
rate stars contribute little to the integrated value.

Independent work with a totally different approach by \citet{Srinivasan08}
obtained a similar integrated mass-loss rate for AGB stars ($2.5
\times10^{-3}$\,\mlu). Our estimate is higher, because 
we included stars with extremely high mass-loss rates, which had not been
categorized before \citep{Blum06}, but were found to be carbon-rich AGB stars
recently \citep{Gruendl08}.

Because of the sensitivity limit in the $\Ks$-band (Fig.\,\ref{Fig-K8-K}),
and possibly in the [3.6]-band (Fig.\,\ref{Fig-3-24}), it is possible that
not all AGB stars with extremely high mass-loss rates will have been
detected.  Therefore the integrated mass-loss rate derived above might be
considered a lower limit.

Table\,\ref{table-lmc-massloss} shows the number distribution per mass-loss
rate range among carbon-rich AGB candidates, combining the two infrared
colour estimates.  Stars with relatively high mass-loss rates
($>6.0\times10^{-6}$\,\mlu) are responsible for most of the integrated
mass loss.  A similar result was found for Galactic AGB stars
\citep{Knapp85, Jura89, LeBertre01, Ojha07}.

There are 1779 carbon-rich AGB candidates in
the LMC (Sec.\ref{selection}) selected by $[3.6]-[8.0]>0.7$.
The number of AGB stars with $[3.6]-[8.0]<0.7$ is almost 10 times greater than
that with $[3.6]-[8.0]>0.7$.  In the region of $10.5<[8.0]<8$ and
$[3.6]-[8.0]<0.7$, there are 13460 stars, where we assume that $[8.0]\sim10.5$
corresponds to the tip of the RGB branch {\it (as the number of stars
  decreases at $[8.0]<10.5$; Fig.\,\ref{Fig-38-8})}, and where both oxygen-rich and carbon-rich
stars are mixed together.  Furthermore, there are carbon-rich AGB stars below
the tip of the RGB ($[8.0]<10.5$) \citep{Lagadec07}, which are uncounted due
to the confusion with RGB stars. These will be a mixture of extrinsic
carbon-stars, which have become carbon-rich due to mass-transfer from an
evolved companion, or AGB stars in the brief phase of low luminosity which
follows a helium shell flash.  Extrinsic carbon stars should not be included
in our census. The numbers of the others could be significant, but their
mass-loss rates are low.  The boundary at $[3.6]-[8.0]\sim0.5$ may be
separating stars with dust-driven winds from those with pulsation/opacity
driven mass loss.  If we assume a plausible mass-loss rate of
$1\times10^{-8}$\,\mlu\, on average, similar to red-giant branch stars
($10^{-9}$--$10^{-8}$\,\mlu\,), the integrated mass-loss rate from these stars
would be $\sim1\times10^{-4}$\,\mlu, a minor contribution to the integrated
mass-loss rate.  These are very rough estimates, and we do not include them
into the final number.

\subsection{Spatial distribution of carbon-rich stars} \label{distribution}

%_________________________________________________________________
\begin{figure}
\centering
%
%\rotatebox{270}{ 
\begin{minipage} {8cm}
%\resizebox{\hsize}{!}{\includegraphics*[107, 23][474, 300]{mass_loss_high.eps}}
\resizebox{\hsize}{!}{\includegraphics*[120, 272][524, 570]{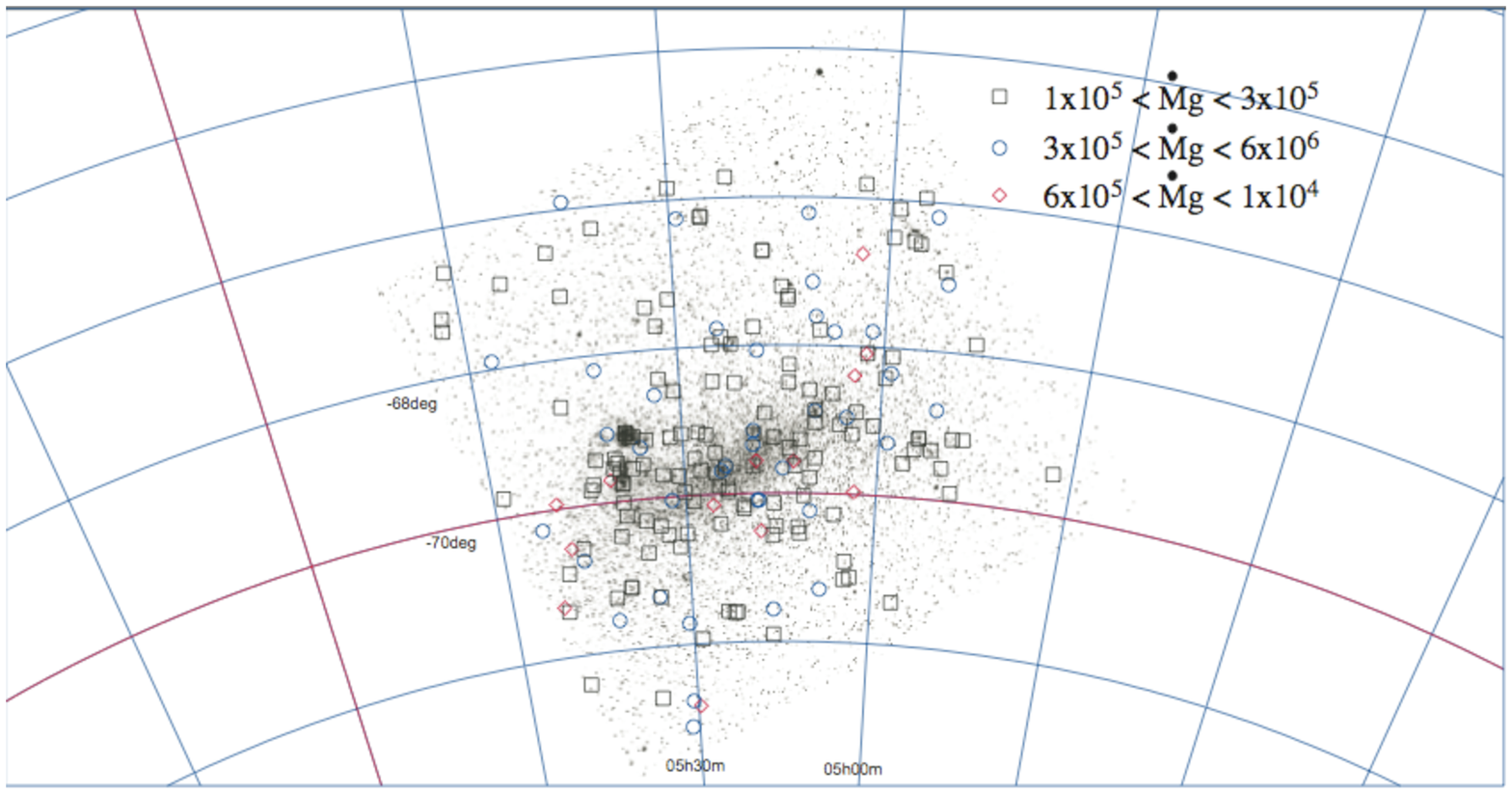}}
\end{minipage}
%}
\vspace{0.2cm}
%
%\rotatebox{270}{ 
\begin{minipage} {8cm}
%\resizebox{\hsize}{!}{\includegraphics*[107, 23][474, 300]{mass_loss_low.eps}}
\resizebox{\hsize}{!}{\includegraphics*[120, 272][524, 570]{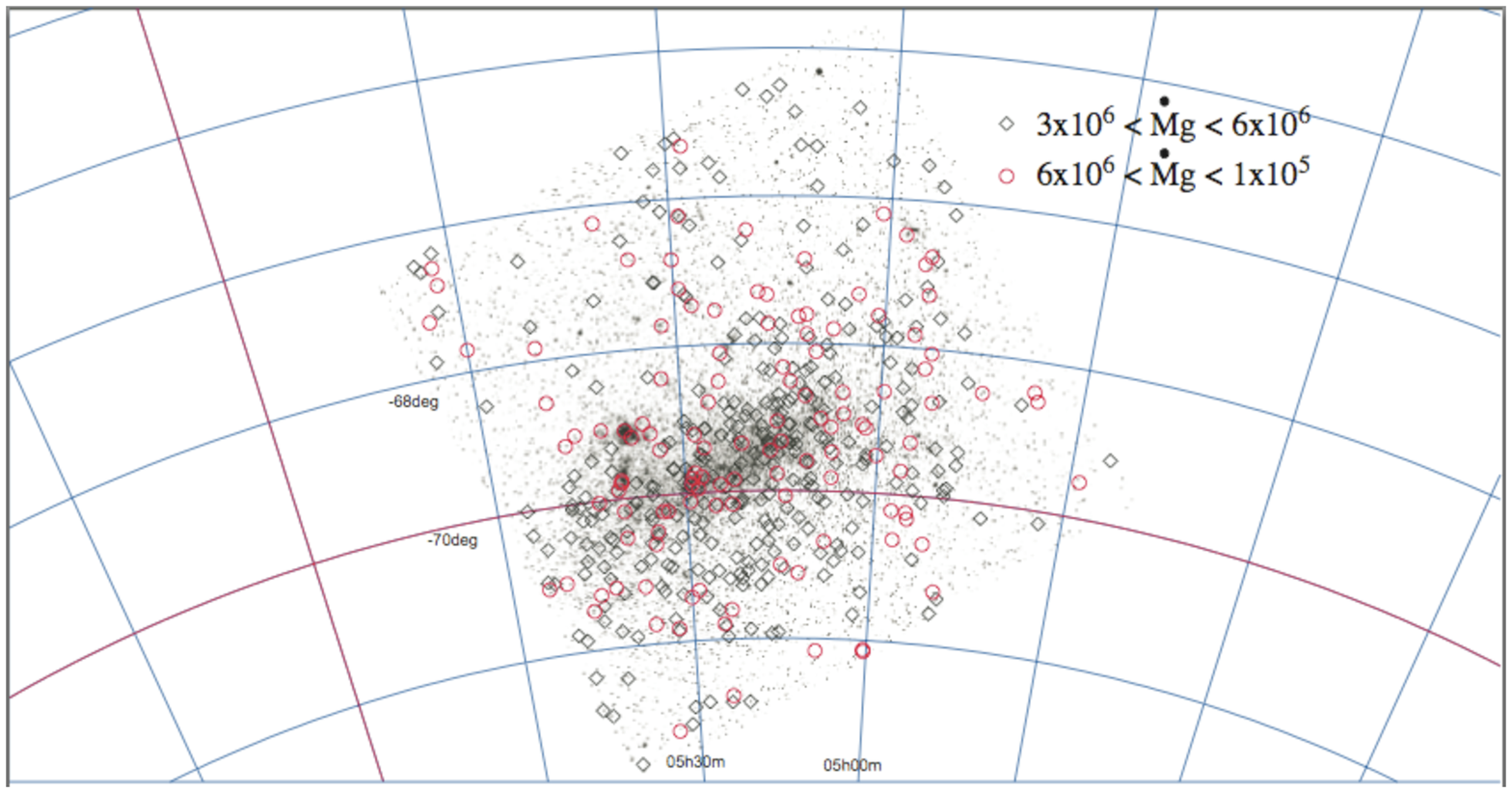}}
\end{minipage}
%}
\vspace{0.2cm}
\rotatebox{270}{ 
\begin{minipage} {6cm}
%\resizebox{\hsize}{!}{\includegraphics*[107, 23][413, 300]{mass_loss_region.eps}}
%\resizebox{\hsize}{!}{\includegraphics*[107, 23][413, 300]{mass_loss_region_color.eps}}
\resizebox{\hsize}{!}{\includegraphics*[141, 253][427, 634]{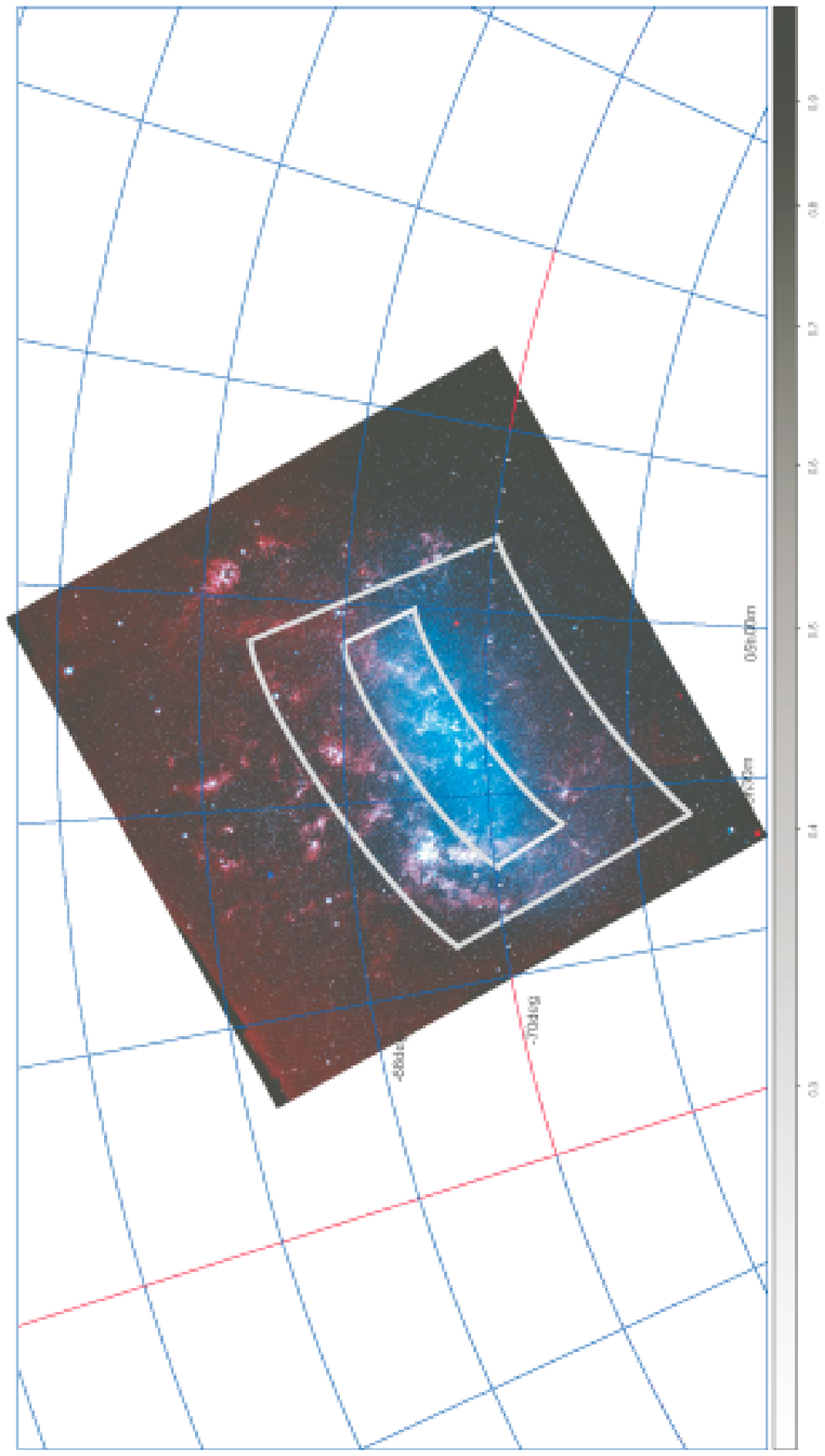}}
\end{minipage}}
\caption{The spatial distribution of mass-losing carbon-rich AGB candidates in
  the LMC is plotted over the SAGE 3.6\,$\mu$m image (top and middle).
  Candidates are classified into groups according to mass-loss rate as given
  on the figure.  Carbon-rich AGB candidates are concentrated in the LMC bar.
  In the bottom panel, large and small rectangles show the areas defined for
  the `bar' and the `bar centre', as used in our analysis.  (on-line version,
  the bottom panel is the SAGE 3-color composition image \citep{Meixner06}:
  3.6\,$\mu$m image for blue, 4.5\,$\mu$m image for green and 5.8\,$\mu$m for
  red.  The LMC bar region is found in blue color, as stellar components are
  dominant.)}
\label{Fig-sp-dist-lmc}
\end{figure}
%_________________________________________________________________

Fig.\,\ref{Fig-sp-dist-lmc} and Table\,\ref{table-sp-dist-lmc} show the
distribution of carbon-rich AGB stars with intermediate and high mass-loss
rates within various parts of the LMC. The areas are specified in the lower
panel of Fig.\,\ref{Fig-sp-dist-lmc}.  The majority of the stars are located
towards the LMC bar.

One-third of the mass-losing carbon-rich AGB candidates are concentrated in
the bar centre.  This is consistent with the findings from near-infrared
surveys, which show a high concentration of carbon-rich stars in the bar
\citep{Blanco83, Cioni06a}.

There are at least two regions containing clusters, NGC\,2154 (RA=05h57m38s,
Dec=$-67^{\rm o}15.7'$, J2000), where about 44 carbon-rich stars have been
reported \citep{Blanco83}, and NGC\,1978 (RA=05h27m17s, Dec=$-70^{\rm
  o}44.1'$, J2000), where about 24 carbon stars have been found by
\citet{Westerlund91}, and which contains a carbon-rich AGB Mira variable
\citep{Tanabe97}.

\section{Discussion}

\subsection{Mass-loss rates from AGB stars per unit area}

Using the result listed in Table\,\ref{table-sp-dist-lmc-prop} for the LMC the
integrated mass-loss rate is $2.2\times10^{-4}$\,\mlu\,kpc$^{-2}$ for
carbon-rich AGB candidates only.  
There is a spatial variation in the range
1.1--5.1\,$\times10^{-4}$\,\mlu\,kpc$^{-2}$.  This value is comparable to that
found in the solar neighbourhood, where \citet{Jura89} investigated AGB stars
within $\sim$1\,kpc from the Sun and obtained an integrated mass-loss rate
from AGB stars of 3--6$\times10^{-4}$\,\mlu\,kpc$^{-2}$. Their study was
limited to stars with \gmas$>2\times10^{-6}$\,\mlu, but lower mass-loss rate
stars would not contribute to the value significantly. Half of the mass was
from oxygen-rich and the other half from carbon-rich AGB stars.
\citet{Thronson87}, \citet{Tielens90} and \citet{Groenewegen92} obtained
similar numbers for the gas injection rate from carbon-rich AGB stars in the
Galaxy. Assuming that the gas-to-dust ratios are comparable in these two
galaxies, the gas mass per unit area injected from carbon-rich stars in the
Galactic plane and the LMC bar centre are similar.

%The stellar mass of the Galaxy and the LMC is approximately 
%$5\times10^{10}$ \citep{Hammer07} and $4\times10^{9}$ \citep{Weinberg99}.

\subsection{Other gas and dust sources in the LMC}

For comparison with the result discussed above, we estimate the gas and dust
mass expelled from other sources in the LMC and  summarize the findings in
Table\,\ref{Table-object-dist}.

%_________________________________________________________________
\begin{table*}
% \centering
  \caption{ Gas and dust injected into the ISM of the LMC}
\begin{center}
 \begin{tabular}{lrrrrrrrrccccccccc}
  \hline
Sources & Gas mass & Dust mass & Chemical type &  \\
 & ($10^{-3}$\,\mlu)& ($10^{-6}$\,\mlu)& \\ \hline
AGB stars \\
~~Carbon-rich                               & 8.6 (up to 20.6$\S$) & 43 (up to 100) & C-rich\\
~~Oxygen-rich                               & $>>$2$\dag$   & $>>$0.4$\dag$ & O-rich \\
Type II SNe                                               &  20--40 & 0.1--130$\ddag$~ & both O- and C-rich \\
WR stars                                       & 0.6 & & (C-rich?)\\
%OB stars                                        & 200--2000? \\
Red supergiants                         & $>$1                 & $>$2~~ & O-rich \\
\hline
\end{tabular}
\label{Table-object-dist}
\end{center}

$\S$ Considering the unknown dust condensation temperature, mass-loss rate can 
be underestimated
by a factor of up to  2.4 (Sect\,\ref{colour-mass-loss}) \\
$\dag$ The number listed is the lower limit obtained from the observations.
Expected gas and dust mass-loss rate are $8.6\times10^{-3}$\,\mlu and $1.2\times10^{-5}$\,\mlu, respectively. \\
$\ddag$ Dust production (or possibly destruction) in and around SNe remains 
uncertain.
% \gmas : gas mass-loss rates estimated using  $\Ks-[8.0]$ and $[3.6]-[8.0]$
\end{table*}
%________________________________________________________________

\subsubsection{Gas and dust injection from SNe}

We estimate the gas output from SNe based on the theoretical model of
\citet{Kodama97}. We use the revised IMF from \citet{Kroupa01} (lower mass
limit: $M_{\rm lo}=0.1\,\rm M_{\odot}$ and upper mass limit: $M_{\rm
  up}=60\,\rm M_{\odot}$), and assume a constant star formation rate (SFR)
over the past few Myr which is the same as the current value. Under these
conditions, the ratio of the SFR to the gas expulsion rate (GER) by type II
SNe is 0.32.  \citet{Smecker-Hane02} measured the SFR using red giants, which
date from as recently as 200\,Myr ago; we adopt this as the current value.
They gave SFRs of 0.01\,\mlu\,deg$^{-2}$ and 0.004\,\mlu\,deg$^{-2}$ in the
bar and disk regions, respectively.  These numbers result in SN gas inputs to
the ISM of $3\times10^{-3}$\,\mlu\,deg$^{-2}$ in the bar and
$1\times10^{-4}$\,\mlu\,deg$^{-2}$ in the disk.  This implies that, in the
bar, the gas injection rate from SNe is higher than that from AGB stars, but
AGB stars are more important in the disk, if oxygen-rich stars are counted.
The fact that supernova remnants (SNRs) are concentrated around the bar region
and 30\,Dor \citep{Williams99} supports this conclusion.

In the following we use the SN rate estimated from the number of SNRs and the
dust mass observed from SNe and SNRs.  \citet{Mathewson83} list 25 SNRs in the
LMC.  They estimated the frequency of type II SNe as one per 275\,yrs, and for
type I SNe as approximately one per 550\,yrs.  More recently,
\citet{Williams99} added a further 6 SNRs.  Although their SN type is unknown,
they are more likely to be Type II, because galaxy chemical evolution models
suggest that Type Ia and Type II occur with a ratio of about 0.3 in the LMC
\citep{Tsujimoto95}.  \citet{Filipovic98} estimated an LMC SNR rate of one in
every 100 yrs.  The Type II SN rate (for both type I and II) of
\citet{Mathewson83} remains valid, but the time interval could be shorter by
up to 145\,yrs.

The dust output from SNe remains uncertain. The famous type II SN in the LMC,
SN\,1987A, emitted a significant infrared excess from dust grains
\citep{Moseley89, Whitelock89}.  The highest estimate of the dust mass from
SN\,1987A is $1.3\times10^{-3}$\,$M_{\odot}$ \citep{Ercolano07}.  Based on
this value, type II SNe could produce up to $5\times10^{-6}$\,\mlu\, of new
dust.  In the SMC dust mass of $8\times10^{-4}$\,$M_{\odot}$ is found in a SNR
\citep{Stanimirovic05}.  Even higher dust masses have been reported from other
type II SNe in other galaxies: 0.02\,$M_{\odot}$ in SN 2003gd in NGC\,628
\citep{Sugerman06} and 0.02--0.05\,$M_{\odot}$ in Cas\,A \citep{Rho08}.  A
lower dust mass has been suggested for SN\,2003gd
\citep[$4\times10^{-5}$\,$M_{\odot}$; ][]{Meikle07}.  \citet{Sakon08}
estimated a dust mass of 2.2--3.4$\times10^{-3}$\,$M_{\odot}$ produced from SN
2006jc.  These estimates give a range of (0.1--130)$\times10^{-4}$\,\mlu\,for
the dust mass from observations of SNe and SNRs.  The dust mass produced by
SNe remains very uncertain for the following reasons: extinction measurements
give a column density rather than a mass; the mid-infrared photometric
measurements are mainly sensitive to warm dust at the early phase, and may
need to be corrected for flash-heated interstellar dust; sub-mm measurements
probe cold dust, but can be confused with interstellar dust clouds. The
clumping is also uncertain and affects the mass determination. Finally, the
primary dust production by SNe may be significantly counteracted by subsequent
dust destruction when the ejecta hits the ISM \citep{Jones94, Dwek08}, which
is not considered here.  The dust production given here for SNe is tentative,
and could change by an order of magnitude or more.

We adopt a lower limit for the progenitor mass of Type II SNe of 
8\,$M_{\odot}$ \citep{Smartt08}, and the IMF from \citet{Kroupa01}, 
with a GER/SFR through SNe of 0.32.  Then, with a SN rate of one per 145--275
yrs, the gas ejection rate from Type II SNe is 0.02--0.04\,\mlu.  This is
higher than the integrated gas mass-loss rate from carbon-rich AGB candidates.
This number critically depends on the progenitor mass range of SNe \citep[see
discussion in][]{Smartt08}. Type Ib/c SNe may also contribute to core-collapse
SNe and are not counted here as their origin remains uncertain.

\subsubsection{WR stars and LBVs}
\citet{Breysacher99} published a catalogue of Population-I Wolf-Rayet (WR)
stars in the LMC.  Among 134 stars, gas mass-loss rates are given for 40, most
of which were measured by \citet{Crowther97}.  Simply integrating these 40
stars gives $2\times10^{-3}$\,\mlu.  If the remaining 74 stars have similar
mass-loss rates on average, we expect about $6\times10^{-3}$\,\mlu\, from
Population-I WR stars in total.  There are over a hundred suspected WR stars
in the LMC, so the gas mass-loss rate from WR stars may be underestimated by a
factor of up to a few. Late-type WR stars (WC7--WC9) produce dust, but no such
stars have been found in the LMC \citep{Smith88, Moffat91}.

Luminous Blue Variables (LBVs) could also potentially contribute to the gas
and dust injected into the ISM.  \citet{Zickgraf06} listed 15 B[e] supergiants
in the LMC and \citet{vanGenderen01} listed 21 S Dor variables.  However, the
mass-loss rates of these stars are still unknown.  Well studied objects of
this type in the Galaxy include $\eta$\,Car and AG\,Car. \citet{Hillier01}
suggested a gas mass-loss rate of $1\times10^{-3}$\,\mlu\, for $\eta$\,Car
using a formula from \citet{Wright75}; near-infrared interferometric
observations suggested $1.6\times10^{-3}$\,\mlu\, \citep{vanBoekel03} for
$\eta$\,Car. \citet{Voors00} estimated a dust mass-loss rate of
$3\times10^{-5}$\,\mlu\, for AG\,Car from mid- and far-infrared observations.
However, it is clear that not all LBVs have dust mass-loss rates as high
as $\eta$\,Car and fewer than half of the objects have noticeable excesses
\citep{vanGenderen01}.

 Winds from OB-type stars could contribute to the gas enrichment process
  in the ISM.  \citet{Bastian09} estimated about 2000 OB stars to be present
  within the inner 3 degree radius of the LMC.  This area is similar in
  size to the SAGE survey coverage.  If  OB stars have mass-loss rates up to
  $1\times10^{-7}$--$1\times10^{-6}$\,\mlu\, \citep{Evans04}, the total gas lost
  from OB stars could be $2\times10^{-4}$--$2\times10^{-3}$\,\mlu.

\subsubsection{Red supergiants} \label{RSGs} High mass stars (more than
8\,\msun) can produce dust grains during their red supergiant phase
\citep[e.g.][]{Verhoelst09}.  One of the most luminous red supergiants (RSGs)
in the LMC is WOH\,G64 and its circumstellar dust mass could be as high as
$2\times10^{-2}$\,$M_{\odot}$ \citep{Ohnaka08}.  This exceeds the total dust
production from SN\,1987A.  \citet{vanLoon05} listed 14 oxygen-rich red
supergiants, which exceed the classical upper limit of AGB luminosity, i.e.
$\log L(L_{\rm \odot})=4.74$ \citep[c.f.][]{Wood83}.  The combined gas
mass-loss rates of these 14 stars is $1\times10^{-3}$\,\mlu.  A gas-to-dust
ratio of 500 was assumed in their work \citep{vanLoon99}.  The dust mass-loss
rate from RSGs amounts to $2\times10^{-6}$\,\mlu.

Previous studies mainly depend on the IRAS survey. After an investigation of
RSGs using SAGE data (Kemper et al. in preparation), the number of
known RSGs may increase. The integrated mass-loss rate for the RSGs quoted
here is therefore likely to be a lower limit.

\subsubsection{Oxygen-rich AGB stars}
The total number of oxygen-rich AGB stars with circumstellar envelopes is
unknown. Thus, their total mass-loss rate remains essentially unknown.
\citet{vanLoon05} discuss 7 oxygen-rich AGB stars and from these we obtain an
integrated mass-loss rate of $2\times10^{-4}$\,\mlu\, in gas and
$0.4\times10^{-6}$\,\mlu\, in dust, assuming a gas-to-dust ratio of 500 as in
the original work \citep{vanLoon99}.  The gas-to-dust ratio of oxygen-rich AGB
stars is unknown, but as silicon atoms are not synthesised in AGB stars, the
silicate dust amount can be assumed to correlate with the original silicate
abundance of the star. It is expected that the gas-to-dust ratios of
oxygen-rich AGB stars are lower in the LMC than in our Galaxy, as a similar
trend was found in the ISM. These provide a lower limit to the integrated
mass-loss rates; further estimates can be made after the IRS followup of SAGE
is completed (Kemper in preparation).

We expect that the upper limits of gas and dust input from oxygen-rich stars
are comparable to the values from carbon-rich stars, simply from the relative
numbers of objects.  It is known that the number ratio of oxygen-rich stars
and carbon-rich stars depends on the metallicities \citep[e.g.]{Feast89,
  Feast06, Groenewegen07a} In the LMC, among AGB stars with low mass-loss
rates there are as many oxygen-rich as carbon-rich stars \citep{Blanco78}.
However, among PNe, the next evolutionary phase, there are only two
oxygen-rich dusty sources found out of 23 PNe \citep[]{Stanghellini07,
  Bernard-Salas08}. Thus, the integrated mass-loss rate from oxygen-rich AGB
stars is expected to be similar to or less than that from carbon-rich stars.
Further, stellar evolutionary models predict that oxygen-rich stars evolve
from both low-mass (1.0--1.5\,$M_{\odot}$ on the main sequence) and
intermediate mass (5.0--8.0\,$M_{\odot}$) progenitors, while carbon stars
originate from 1.5--5.0\,$M_{\odot}$ at the LMC metallicity \citep{Karakas07}.
Many intermediate-mass oxygen-rich stars will have experienced hot bottom
burning \citep{Boothroyd93}, turning carbon atoms into oxygen atoms.
\citet{Kroupa01}'s IMF gives an oxygen- and carbon-rich ratio of 1:1 in
number, but 1:1.4 in mass.  Hence our conclusion is that the integrated
mass-loss rate from oxygen-rich AGB stars is expected to be similar to or less
(approximately $6\times10^{-3}\,\msun$) than that from carbon-rich AGB stars.

\citet{LZ08} argue that at low metallicity, 
oxygen-rich dust drives a superwind only when the stellar luminosity
exceeds a crictical value, of the order of $10^4\,\rm L_\odot$ for the case of
the LMC. This would predict that the mass loss is more dominated by carbon
stars than their number ratio would suggest.

 Among AGB stars, S-type stars represent stars with a C/O ratio close to
  unity.  This type of star are considered to trace a transiting phase from
  oxygen-rich stars to carbon-rich stars.  In the LMC, there is no complete
  sensus of S-type stars; a few stars are listed in \citet{Cioni01}.  In
  our Galaxy, the fraction of S-type stars is assumed to be 1\,\% of total AGB
  population.  Thus, we assume gas and dust lost from S-type stars is
  relatively minor.  Except for a few SC-type stars \citep{Aoki98}, which have
  a  carbon-rich atmosphere, S-type stars tend to produce some silicate dust
  \citep{Little-Marenin88, Chen93}.

\subsection{Gas and dust sources in the LMC}

Table\,\ref{Table-object-dist} summarizes the estimated gas and dust
production from various sources.  AGB stars are one of the main sources of
the dust enrichment for the ISM of the LMC, and carbon-rich AGB stars are a
major factor.  The dust contribution from SNe is very uncertain. There could
also be minor inputs from novae and R\,CrB stars. Dust evolution models show
that within our Galaxy carbon dust originates mainly from carbon-rich AGB
stars \citep{Dwek98, Zhukovska08}. Our work shows that dust production in
the LMC follows a similar trend.

If a gas-to-dust ratio of 200 is valid for carbon-rich AGB stars, and our
estimates for mass loss from SNe are correct,  AGB stars are as important
  a gas contributor as are SNe.
SN activity is concentrated
towards the 30\,Dor and LMC bar region.  In the LMC disk, AGB stars could be
the more important source of gas and dust.
Other sources appear to present minor
contributions to the ISM enrichment.

This result is in marked contrast to our Galaxy where AGB stars are thought to
be the main source, even for gas \citep{Tielens05}.   The contribution
  from SNe is an order of magnitude smaller than that of AGB stars.  The 
  observational difference can be related to the recent increase in the star
formation rate \citep[SFR; ][]{Smecker-Hane02} in the LMC bar region, which
has been suggested to be due to the tidal interaction with the SMC
\citep{Bekki05}. Indeed, assuming a constant SFR and using the IMF, we find
that the gas injection rate from AGB stars should be a factor of twice larger
than that from SNe and their precursors. Thus an increase of the SFR appears
to be required in the LMC. 

\subsection{Populations and spatial distributions} \label{population}
 
Table\,\ref{table-sp-dist-lmc} shows, as a function of mass-loss rate,
the number distributions for carbon-rich AGB candidates in the bar, the bar
centre, outside of the bar, and the entire area surveyed.  We have defined
the bar centre and bar regions as shown in Fig.\,\ref{Fig-sp-dist-lmc}. The
fraction of stars with different mass-loss rates shows little dependence on
their location within the LMC.  This suggests that most of these stars
belong to a similar population.  Contamination by non-LMC objects, i.e.,
distant galaxies, should be minimal; the number density of distant galaxies
is almost uniform and thus the number counted should be proportional to the
area examined.  

\citet{Reid06} surveyed PNe in the LMC using narrow-band optical filters.
They found fewer PNe in the north part of the bar (between 5h45m, $-68^{\rm
  o}30'$ and 5h15m, $-67^{\rm o}15'$) than in the surrounding area, and
concluded that interstellar extinction towards this region is responsible for
the lower number density.  However, a similar trend is found for the AGB-star
distribution, despite the fact that extinction in the near- and mid-infrared
will be negligible.  Generally, in comparison to the age of galaxies, AGB
stars and PNe belong to similar stellar populations. Thus, this absence of PNe
and AGB stars in the north part of the bar appears to be intrinsic to the LMC.

A distinct component of the population is also found among young red
supergiants; fig.~2 in \citet{Cioni00a} shows a secondary peak at about $\rm
Dec>-67^o$ which presumably corresponds to stars younger than 0.5\,Gyrs.
Similarly, high mass-loss rate AGB stars (rates between
$6\times10^{-6}$\,\mlu\, and $3\times10^{-4}$\,\mlu) are populous in that
region, although they are absent in the northern part of the bar. In this
region, stars with low and intermediate mass-loss rates (from
$3\times10^{-6}$\,\mlu\, to $6\times10^{-6}$\,\mlu) are under-populated.  This
may suggest that the location of active  star forming regions has not
changed significantly over the past 0.5--1\,Gyrs.

Oxygen-rich AGB stars originate both from relatively low mass
(1.0--1.5\,$M_{\odot}$) and relatively high mass 
(5.0--8.0\,$M_{\odot}$) stars, at the LMC metallicity \citep{Karakas07}.  The
distribution of oxygen-rich and carbon-rich stars might differ from 
each other, as the majority of the oxygen-rich stars (which
belong to a lower mass group), belong to an older population.  Indeed,
\citet{Blanco83} and \citet{Cioni06a} showed that oxygen-rich stars 
spread further out into the disk region than do carbon-rich stars.

Based on optical imaging and a population synthesis model,
\citet{Smecker-Hane02} suggested that in the bar region
35 percent of the stars are younger than 3\,Gyr, and 71 percent are younger
than 7.5\,Gyr.  In contrast, the disk has an older population, as the
corresponding numbers are 19 percent and 41 percent.  The initial masses of
carbon-rich stars are 1.5--5.0\,$M_{\odot}$ \citep{Vassiliadis93,
Stancliffe05}.  These stars reach the thermal pulsating AGB phase in under
$3$\,Gyr.  Our observations show that the carbon-rich AGB candidates are
concentrated in the bar, which is consistent with a population younger than
3\,Gyr in the LMC.

\subsection{Evolution of the ISM}

%_________________________________________________________________
\begin{table*}
% \centering
  \caption{ Evolution of the ISM}
\begin{center}
 \begin{tabular}{lrrrrrrrrccccccccc}
  \hline
{\it Gas budget} \\
Stellar feedback (AGB + SNe) & 0.03 -- 0.07\,\mlu \\
Star formation rate (SFR) & 0.19 -- 0.26\,\mlu \\
Gas consumption rate (AGB + SNe $-$ SFR) &  $-$0.23 -- $-$0.12\,\mlu \\
Existing gas in the ISM (H{\small I} + H$_2$)  & $8\times10^8$\,\msun \\
Time to exhaust ISM gas & $\sim3\times10^9$ years \\
\\
{\it Dust budget} \\
AGB dust production rate &  (5.5--11.5)$\times10^{-5}$\,\mlu \\
SNe dust production rate & (0.01--13)$\times10^{-5}$\,\mlu \\
SNe dust destruction rate & ? \\
Life time of dust & 2--4$\times10^8$ years \\
Accumulated dust mass over the life time of dust (with constant rate) 
 &  $5\times10^4$\,\msun \\
Existing dust in the ISM &  $1.6\times10^{6}$\,\msun \\
Age of the LMC & 10--15 Gyrs \\
\hline
\end{tabular}
\label{Table-ISM-evolution}
\end{center}
\end{table*}
%________________________________________________________________

\subsubsection{Star formation rate and stellar feedback}

The LMC SFR is higher than the gas injection rate from SNe and AGB stars (in
total 0.03--0.07\,$M_{\odot}$\,yr$^{-1}$).  \citet{Whitney08} estimated a SFR
of 0.19\,$M_{\odot}$\,yr$^{-1}$, and \citet{Kennicutt95} suggested a SFR of
0.26\,$M_{\odot}$\,yr$^{-1}$.  These are an order of magnitude higher than the
gas output from AGB candidates and SNe.  This suggests that the current LMC
star formation depends on the large reservoir of ISM gas
($7\times10^8$\,$M_{\odot}$ in H{\small I}; \citet{Westerlund97} and
$1\times10^8$\,$M_{\odot}$ in H$_2$; \citet{Israel97}).  This is summarised in
Table\,\ref{Table-ISM-evolution}.   Without continuing infall, the LMC
  would become gas poorer, due to the excess SFR, and the SFR in the
  LMC would eventually decline.  The Magellanic stream has in fact removed
  further material from the LMC.  Further, stellar feed back for at least
  1\,Gyrs is required to gain 10\% of gas mass present in the ISM, suggesting
  stellar yields can impact the ISM abundance only over a time scale of a few
  Gyrs.  This is consistent with a time scale of the metallicity increases
  (order of a Gyr), according to the age-and-metallicity relation
  \cite[e.g.][]{Feast89, Russell92, Pagel98, Carrera08}.

\subsubsection{Impact on extinction curve}

The main sources of gas and dust are different in our Galaxy and in the LMC.
This begs two questions: 1) why does the gas-to-dust ratio in the ISM
in the LMC differ from that in the Galaxy? 2) Why do the proportions of
amorphous silicate and carbon dust grains in AGB stars appear to differ
from those in dust grains in the ISM.

  The different enrichment may be expected to have an impact on the ISM.
  \citet{Gordon03} measured that the gas-to-dust ratio varies within the LMC.
  The highest values was found in the LMC `supershell' (N(H{\small
    I})/$A(V)$=5--23$\times10^{21}$\,atoms\,cm$^{-2}$\,mag$^{-1}$), but
  overall gas-to-dust ratio remains similar (N(H{\small
    I})/$A(V)$=1--7$\times10^{21}$\,atoms\,cm$^{-2}$\,mag$^{-1}$) but slightly
  higher than the Galacitic value (N(H{\small
    I})/$A(V)$=1.5$\times10^{21}$\,atoms\,cm$^{-2}$\,mag$^{-1}$).
  \citet{Gordon03} explained the variation with the local condition.  The LMC
  `supershell' is strongly affected by SNe, destroying dust grains.

  The extinction curve in the LMC is quite similar to the Galactic one, except
  for a slight steep rise in far-UV region in 30\,Dor region \citep{Gordon03}.
  The 2175\,\AA\, bump is assumed to be associated with carbonaceous dust
  grains; this feature is weak in the 30\,Dor region.  Although the LMC has a
  higher carbonaceous dust input for AGB stars, compared to silicate dust,
  than does our Galaxy, this does not appear in the extinction curve.  As
  \citet{Gordon03} suggested, the extinction curve in 30\,Dor region is
  strongly affected by SN dust destruction. However, overall the consistency
  of the extinction curve in the LMC and our Galaxy can not be explained with
  SN dust destruction. An additional silicate dust source may be required.

\subsubsection{AGB feedback with respect to past SFR in the LMC and our Galaxy}

We can estimate the gas and dust feedback rate from carbon-rich AGB stars and
compare it with the mass removed from the ISM by star formation. We adopt the
SFR in the bar and the disk from \citet{Smecker-Hane02} of
0.01\,\mlu\,deg$^{-2}$ and 0.004\,\mlu\,deg$^{-2}$, respectively. These
numbers approximately trace the SFR 1--3 Gyr ago, when most carbon-rich AGB
stars were formed. The gas ejection rate (GER$_{\rm cAGB}$) and dust ejection
rate (DER$_{\rm cAGB}$) from carbon-rich AGB stars are taken from
Table\,\ref{table-sp-dist-lmc}. In the bar area, GER$_{\rm cAGB}$/SFR is
0.03--0.04 (up to 0.04) in mass, and outside of the bar the ratio is 0.02,
i.e., the two regions have similar rates.  This ratio could depend on the IMF,
but no apparent difference is found in these two GER$_{\rm cAGB}$/SFR. In the
LMC, more high mass-loss rate AGB stars tend to be carbon-rich than
oxygen-rich.

If we include the oxygen-rich stars, GER$_{\rm AGB}$/SFR could be higher than
GER$_{\rm cAGB}$/SFR by a factor of 1.4--2, resulting in GER$_{\rm AGB}$/SFR
of 0.03--0.08.   If the SFR is approximately constant over the life time
  of the AGB stars, about 3--8\,\% of the stellar mass is returned to the ISM
  as gas through AGB stars. The uncertainties do not include the effect of the
  unknown condensation temperature of carbon-rich dust grains, which results
  in a systematic error of up to a factor of two on the colour and mass-loss
  rate relation.  The ratio of the dust ejection rate from carbon-rich AGB
stars to the SFR is 1--$2\times10^{-4}$.

Similarly, we estimate the ratio of the gas ejection rate from AGB stars
(GER$_{\rm AGB}$) to the SFR for Galactic stars, where we obtain 0.02--0.06
for GER$_{\rm AGB}$/SFR and 0.8--3$\times10^{-4}$. Here we adopt the
integrated AGB mass-loss rate of 3--6$\times10^{-4}$\,\mlu\,kpc$^{-2}$
\citep{Jura89}. In this case the sample is not limited to carbon-rich stars,
as oxygen-rich stars are included. The SFR of our Galaxy is very uncertain and
we adopt the value from \citet{Rana91}, who suggest
0.01--0.02\,\mlu\,kpc$^{-2}$ at 3\,Gyrs ago. The measured AGB feedback rates
in the LMC and the Galaxy are similar.

 The AGB feedback rate can be compared with theoretical estimates.  Stars
within the mass range 1.0--8.0\,$M_{\odot}$ comprise about 24 percent of the
total mass of stars, according to \citet{Kroupa01}. If these low- and
  intermediate-mass stars lose about 50--80 percent of their mass during the
AGB phase, as estimated from the white dwarf mass, then 12--19 percent
of the feedback to the ISM should theoretically be from these stars.   The
  AGB feedback rate is only 3--8 percent of the stellar mass, with a potential
  increase up to 16 percent.  In general, our observations are consistent with
  stellar feedback through AGB stars but it appears that the theoretical
  stellar feedback is slightly higher than the observed one.

\subsubsection{Missing dust-source problem}\label{missing-dust}

 The total dust mass in the ISM is approximately $1.6\times10^{6}$\,\msun.
  For this estimate, we use the total ISM gas mass in the LMC of
  $8\times10^{8}$\,\msun, and assume that the dust to gas extinction ratio in
  the LMC is about a factor of 2--2.5 lower than the Milky Way
  \citep{Gordon03, Cox06}, the adopted gas-to-dust ratio, of 500 in the LMC,
  is higher than the Galactic value \citep{Zubko04}.

  The total dust injection rate from carbon stars is measured to be
  $4.3\times10^{-5}$\,\mlu\, and up to $1.0\times10^{-4}$\,\mlu.  The rate
  from oxygen-rich stars is expected to be $1.2\times10^{-5}$\,\mlu.  This is
  summarised in Table\,\ref{Table-ISM-evolution}.  Ignoring dust destruction,
  the dust mass in the ISM corresponds to about 20\,Gyrs at the current
  injection rate, which is far too long. In fact, the life time of dust grains
  is of the order of 2--4$\times10^8$\,years, due to destruction by SN shocks
  \citep{Jones94}, rather less than 20\, Gyrs.  Even adopting the highest dust
  production rate from SNe, the existing dust mass ($8\times10^{8}$\,\msun) in
  the ISM is an order of magnitude high than the expected dust mass
  ($5\times10^4$\,\msun) over constant dust supply from SNe and AGB stars over
  $4\times10^8$years.  This shows a `missing dust-source problem' in the LMC.

  A similar missing dust source problem is also found in high-z galaxies
  \citep{Hughes98, Morgan03}.  \citep{Meikle07} found that the total dust
  produced from SNe should be far less than the dust detected in the high-z
  galaxies \citep[e.g. $10^8$\,\msun; ][]{Bertoldi03}.  Recently
  \citet{Sloan09} found that AGB stars can contribute to dust production in
  addition to SNe, even for galaxies at z$\sim$6.  But although the dust
  production AGB stars can alleviate the discrepancy, it is unlikely to
  resolve it in high-z galaxies.  Additional dust sources are therefore needed
  in the ISM in both the LMC and high-z galaxies.  One possibility is star
  formation regions, as suggested by \citet{Morgan03}.  Alternatively, dust
  processing and dust growth by gas accretion in the ISM may yield a total
  dust mass in the ISM which exceeds the dust mass produced by stars.

\section{Conclusions}

We have analysed Spitzer data for AGB stars in the LMC, and presented a
quantitative analysis of gas and dust budge in the LMC ISM.  We obtain the
following main conclusions:
\begin{itemize} 
\item The mass-loss rates of the AGB stars are a quantifiable function of
  their infrared colours.

\item The total mass-loss rate from all carbon-rich AGB candidates is
  estimated to be $8.6\times10^{-3}$\,\mlu\, (up to $21\times10^{-3}$\,\mlu)
  in gas and $4.3\times10^{-5}$\,\mlu\, (up to $1\times10^{-4}$\,\mlu) in dust
  in the LMC.  Adding oxygen-rich AGB stars could increase the gas feedback
  rate to $15\times10^{-3}$\,\mlu\, (up to $27\times10^{-3}$\,\mlu) and the
  dust feedback rate to $6\times10^{-5}$\,\mlu\, (up to
  $1\times10^{-4}$\,\mlu) from AGB stars.

\item AGB stars are among the main sources of gas and dust in the
  LMC.  AGB stars are the dominant source of gas in the disc, but SNe are a
  slightly more important gas source in the bar region of the LMC. The gas
  injection rate from SNe could be as high as $4\times10^{-2}$\,\mlu\, in the
  LMC, but the dust production of SNe is uncertain.

\item The measured feedback from carbon-rich AGB stars into the ISM is 2--4
  percent of the SFR 3 Gyr ago for gas mass, and 0.01--0.02 percent of the SFR
  for dust mass. The gas feedback from both carbon-rich and oxygen-rich AGB
  stars is expected to be 3--8 percent of the SFR. These values could increase
  by a factor of two, due to the uncertainty of dust condensation temperature.
  These ratios appear to be similar in number within the LMC bar, the disk and
  Galactic disk.

\item The gas output of SNe is found to be more important in the LMC bar than
  in both the remaining region of the LMC and in the Galaxy, related to a
  higher gas-to-dust ratio in the LMC.
 
\item The total gas mass supplied through AGB stars and SNe is far less than
  the current star formation rate.  The star formation largely depends on the
  gas already present in the LMC ISM, and without infall, eventually the SFR
  will decline as a consequence of the gas exhaustion in the ISM.

\item In the LMC, there is `a missing dust source' problem present. The
  largest estimate of dust accumulation from AGB stars and SNe can not account
  for the full dust mass in the ISM.  This is a similar problem as found for
  high-z galaxies.  Additional dust sources are required, possibly located in
  star-forming regions.

\end{itemize}

\section{acknowledgments}
We thank an anonymous referee for useful comments and suggestions which helped to improve this paper.
M.M. is a JSPS fellow. A.A.Z., F.K., and E.L. acknowledge a STFC rolling grant which
supported this research. M.M thanks to Prof. N. Arimoto and Mr. D Stock for
inputs about evolution of galaxies, and Miss J. Fabbri for inputs about SN gas
and dust, Prof. A.G.G.M. Tielens for inputs about the Galactic gas and dust
budget, and Drs. N Bastian and C. Evans, and Prof. I.D. Howarth for inputs
about OB stars in the LMC.  The SAGE Project is supported by NASA/Spitzer
grant 1275598 and NASA NAG5-12595.  This research has made use of the SIMBAD
database, operated at CDS, Strasbourg, France.

\appendix
\section{Classification of carbon-rich stars}
 \subsection{Existing classification schemes} \label{other-colours}
 
\begin{figure}
\centering
%\rotatebox{90}{ 
%  \begin{minipage} {6.5cm}
%  \resizebox{\hsize}{!}{\includegraphics*[11,20][523,679]{JK_K_nomark_lo.ps}}
%  \end{minipage}}
\rotatebox{90}{ 
  \begin{minipage} {6.5cm}
%  \resizebox{\hsize}{!}{\includegraphics*[11,20][523,679]{JK_K_lo.ps}}
    \resizebox{\hsize}{!}{\includegraphics*[11,20][523,679]{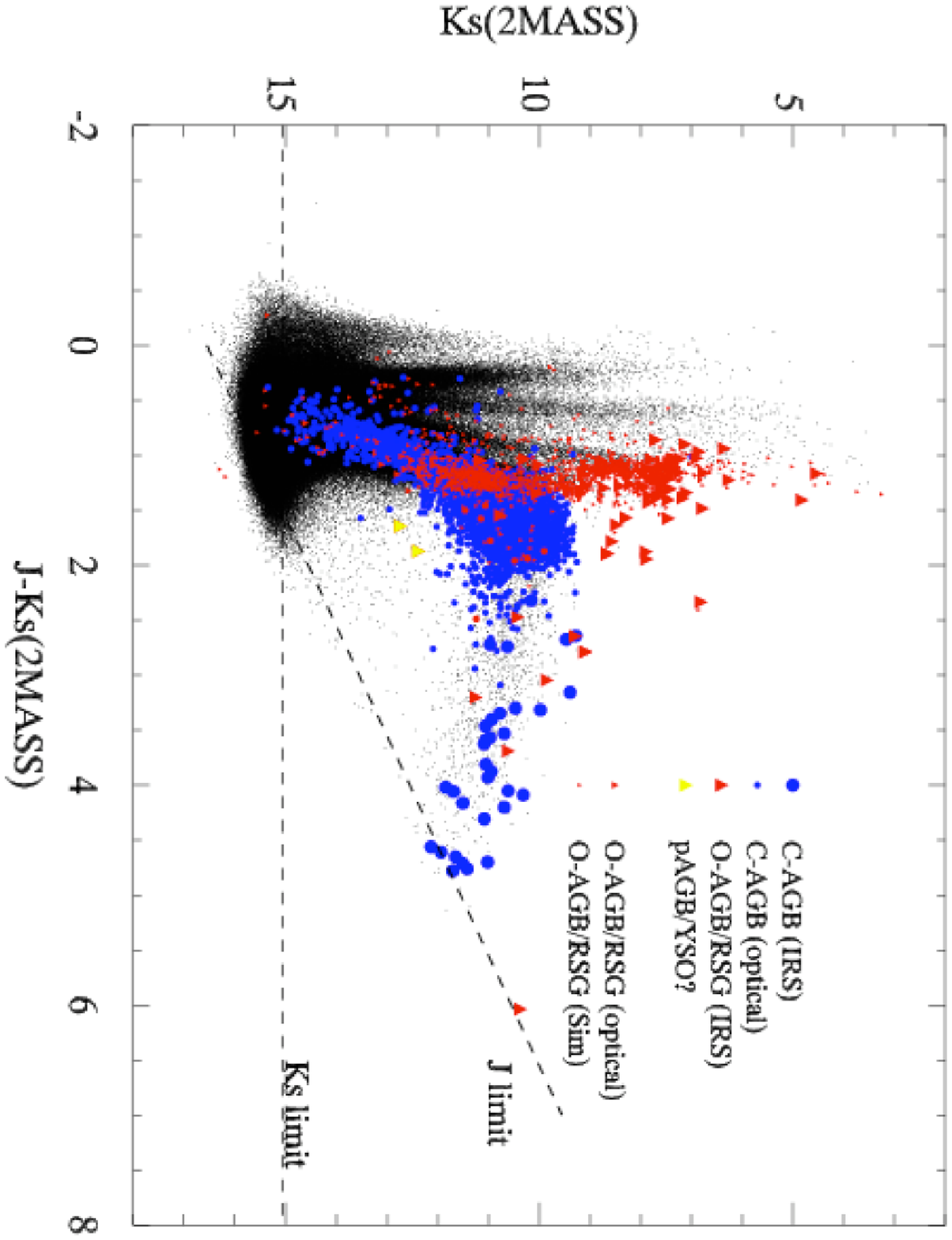}}
  \end{minipage}}
%\resizebox{0.5\hsize}{!}{\includegraphics*[11,106][569,729]{Nikolaev_fig8.ps}}
%\resizebox{0.5\hsize}{!}{\includegraphics*[28,128][519, 730]{Cioni_2003_fig2.ps}}
\caption{
2MASS $J-\Ks$ vs $\Ks$ colour magnitude diagram for the LMC, extracted
from the SAGE catalogue. The lower graph shows cross-identifications with
carbon-rich AGB stars, oxygen-rich AGB stars and RSGs. Large blue circles:
carbon-rich AGB stars according to their IRS spectra. Small blue circles:
carbon-rich AGB stars from \citep{Kontizas01}, classified from optical 
spectra. Red triangles: oxygen-rich red giants and supergiants based on
IRS spectra.  
%No interstellar corrections are adopted.
%(convert into Hess diagramme)
}
\label{Fig-JK-K}
\end{figure}

 Fig.\,\ref{Fig-JK-K} shows a $J-\Ks$ vs $\Ks$ colour magnitude diagram for
 the LMC stars from the SAGE catalogue. The coding of the chemical types of
 the stars is given in the figure. There are five branches within $11>\Ks>7$.
 Carbon-rich stars are found in the reddest of these branches ($J-\Ks
 \gsim1.4$), as is known from previous studies \citep{Nikolaev00}. From $J-\Ks
 \sim1.4$ to 2.0\,mag there is a bend in the carbon-rich branch and the
 $\Ks$-mag becomes brighter with colour.  This branch reaches its brightest
 $\Ks$-mag at about $J-\Ks\sim2$, after which $\Ks$ becomes fainter. The
 number density also decreases redward of $J-\Ks\sim2$. Stars with $J-\Ks>2$
 are described as obscured AGB stars by \citet{Nikolaev00}, and in our sample
 most of these are carbon-rich, but oxygen-rich stars are also found here.
 Following the classification by \citet{Cioni06a} ($\Ks < -0.48\times( J -
 \Ks)+ 13$ and $\Ks < -13.333\times( J - \Ks ) + 24.666$), only 17 percent of
 the stars are redder than $J-\Ks >2$. This shows that less than one-sixth of
 carbon-stars have a circumstellar envelope detectable in the $\Ks$-band. This
 is consistent with the stellar evolution model by \citet{Marigo07}, who
 showed that the super-wind phase (with mass-loss rate higher than
 $5.6\times10^{-7}$\,\mlu) lasts less than 15 percent of the thermal-pulsing
 lifetime for carbon-stars.
%The tip of red giant branch (RGB) is about 12.0\,mag \citep{Cioni00},
%and there are some carbon-rich stars below this criteria.

\begin{figure}
\centering
\rotatebox{90}{ 
\begin{minipage} {6.5cm}
%\resizebox{\hsize}{!}{\includegraphics*[11,20][523,679]{J3_3_lo.ps}}
\resizebox{\hsize}{!}{\includegraphics*[11,20][523,679]{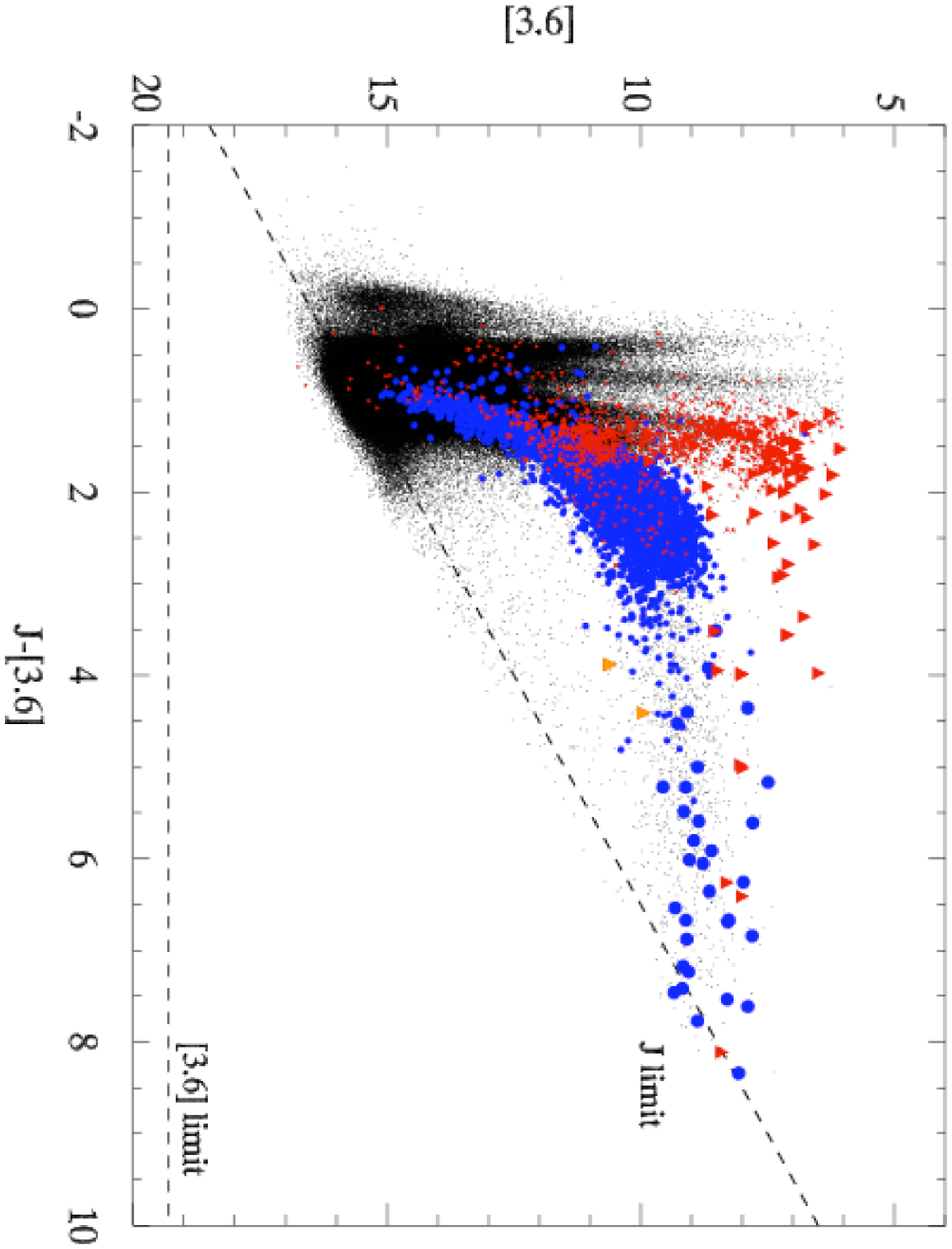}}
\end{minipage}}
\caption{
$J$-band and $J-[3.6]$ colour magnitude diagram.
%$J$-band is the sensitivity limit.
}
\label{Fig-J3-3}
\end{figure}

\begin{figure}
\centering
\rotatebox{90}{ 
\begin{minipage} {6.5cm}
%\resizebox{\hsize}{!}{\includegraphics*[11,20][523,679]{K8_k_nomark_lo.ps}}
\resizebox{\hsize}{!}{\includegraphics*[11,20][523,679]{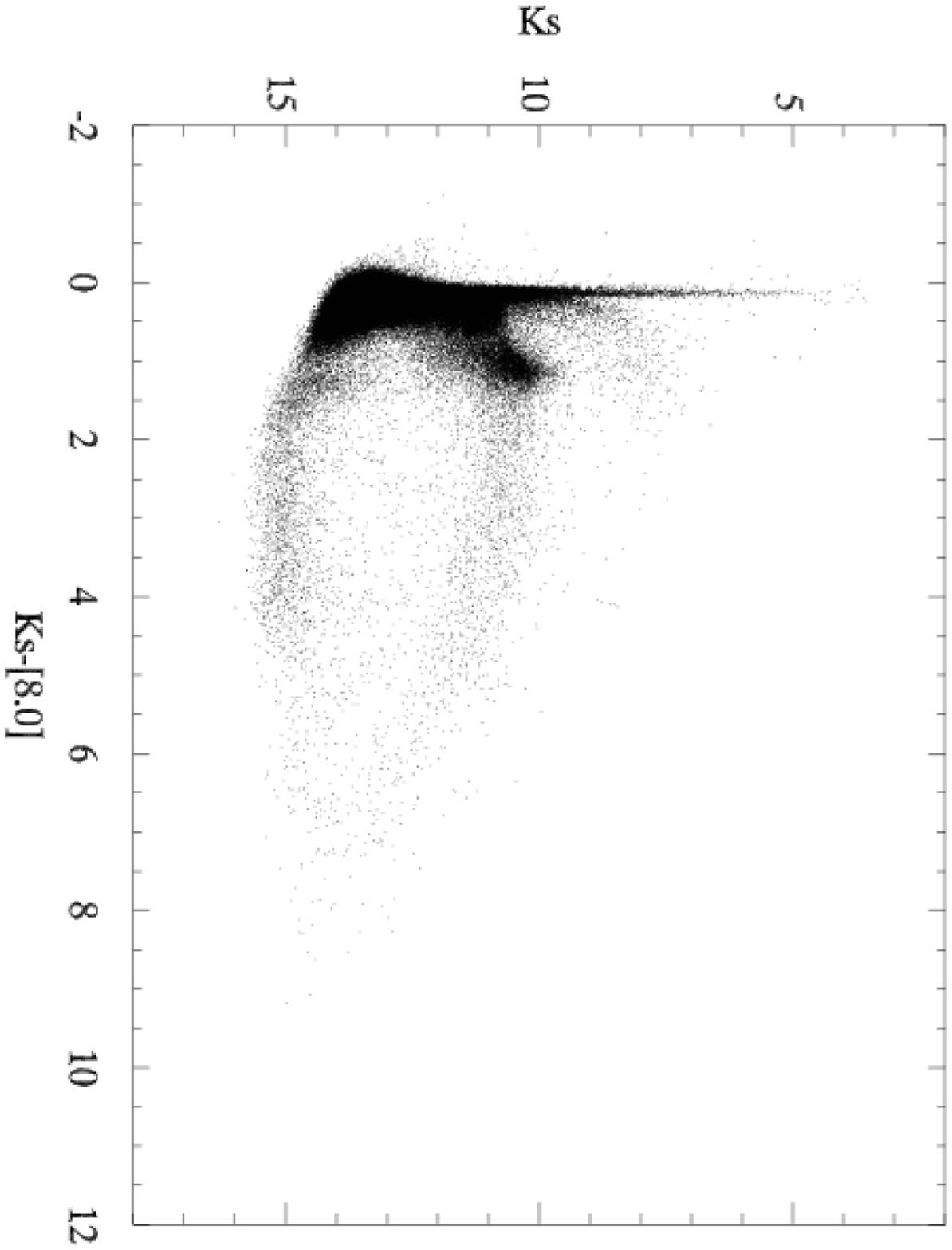}}
\end{minipage}}
\rotatebox{90}{ 
\begin{minipage} {6.5cm}
%\resizebox{\hsize}{!}{\includegraphics*[11,20][523,679]{K8_k_lo.ps}}
\resizebox{\hsize}{!}{\includegraphics*[11,20][523,679]{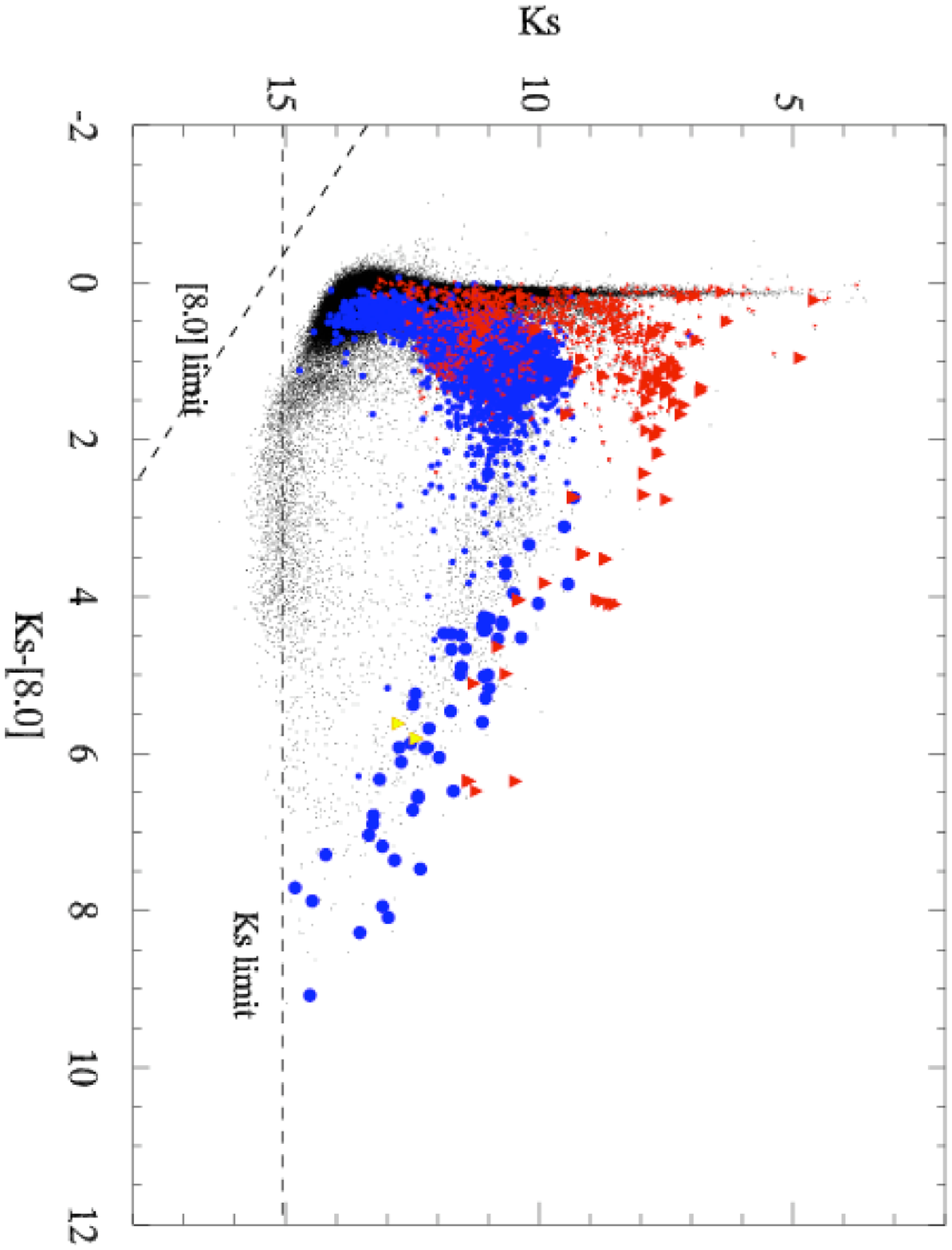}}
%\resizebox{\hsize}{!}{\includegraphics*{K8_k_o_lo.ps}}
\end{minipage}}
\caption{ $\Ks$ and $\Ks-[8.0]$ colour magnitude diagram. The upper panel
shows all of the SAGE sample, and spectroscopically known oxygen-rich
AGB/RSGs, while carbon-rich AGB stars are indicated in the lower panel.
The $Ks$-band sets the detection limit at $\Ks-[8.0]<2$\,mag, except for
blue ($\Ks-[8.0]<2$\,mag) stars, which are [8.0]-band limited. 
The number of carbon-rich AGB stars decreases at about $\Ks-[8.0]$=1.4
%The dotted lines
%show the criteria used to separate the oxygen-rich stars from
%the carbon-rich AGB ones.
}
\label{Fig-K8-K}
\end{figure}

The 2MASS sensitivity is not sufficiently high to detect dusty carbon-rich
AGB stars in some bands. Figs.\,\ref{Fig-J3-3} and \ref{Fig-K8-K} show that
2MASS does not detect the $J$- and $\Ks$-bands at the red end of the
carbon-rich sequence. However, this red end is clearly detected at [5.8] and
[8.0] (Fig.\,\ref{Fig-38-8} and Fig.\,\ref{Fig-58-8}). The situation will
change once the VISTA LMC survey of the Magellanic System is completed,
which could detect as faint as \Ks=20.3\,mag \citep{Cioni08}.

Oxygen-rich stars are mainly found at \Ks$\sim$8\,mag and $J-\Ks<1.5$. These
stars are a mixture of red giants and red-super giants \citep{Kastner08}.
\citet{Cioni03} show that oxygen-rich AGB stars are found in the branch
located at about $\Ks\sim$9\,mag and $J-\Ks<1.5$. In our analysis
(Fig.\,\ref{Fig-JK-K}), there are two regions with oxygen-rich stars. One is
the same as identified by \citet{Cioni03}. The other is occupied by
oxygen-rich stars with think dust shell, which are just above the
obscured carbon-rich sequence. There are two oxygen-rich stars with lower
luminosity than the others ($\Ks\sim$13\,mag; $J-\Ks$=1.5--2.0): l63-31(SAGE
045433.86$-$692036.1) and MSX\,LMX\,610. Their low luminosities and some
excess at near-infrared wavelengths suggest that they could be oxygen-rich
post-AGB candidates or YSO candidates. Post-AGB stars with optical
variability show similar $J-\Ks$ and \Ks-magnitudes
\citep{Wood01}. The possibility that they are YSOs is not discarded, but if so
the $\Ks\sim 13$ mag implies a relatively high mass
\citep[between 5 and 10\,$M_{\odot}$;][]{Whitney08}.

Fig.\,\ref{Fig-J3-3} shows the $J-[3.6]$ vs $[3.6]$ colour magnitude diagram,
used by \citet{Blum06}, in which the separation of oxygen- and carbon-rich AGB
stars is clear. Oxygen-rich stars have OH in the $J$-band, and most Mira
variables have H$_2$O in $JHK$- and $L$-bands \citep{Lancon00, Tsuji97,
  Matsuura02a, Tej03}.  Carbon-rich stars have HCN and C$_2$H$_2$ bands at
3.1\,$\mu$m \citep{ Ridgway78, Groenewegen94, vanLoon99, Matsuura02b,
  Matsuura05} and CN in the $J$- and $H$-bands \citep{Whitelock87, Lancon00}.
The sensitivity is limited by the $J$-band, and heavily dust-obscured AGB
stars are not detected. In Fig.\,\ref{Fig-JK-K} optically identified carbon
stars listed by \citet{Kontizas01} reach the brightest \Ks\,mag.  The \Ks\,
magnitudes become fainter as the stars become redder in $J-\Ks$, as they do
among the more obscured carbon stars classified by Spitzer IRS. In contrast,
such a trend is not found in Fig.\,\ref{Fig-J3-3}; the [3.6] magnitude
continues to brighten among obscured carbon stars beyond the $J$-band
detection limit, as would be expected.

%_________________________________________________________________
%_________________________________________________________________
%_________________________________________________________________
%_________________________________________________________________

%_________________________________________________________________
%_________________________________________________________________

\subsection{Classification using $\Ks-[8.0]$ vs $\Ks$}

Fig.\,\ref{Fig-K8-K} is a colour-magnitude diagram showing $\Ks-[8.0]$ vs
$\Ks$. In this carbon-rich stars become fainter in $\Ks$\, as circumstellar
reddening changes the colour; $\Ks$\, reaches a peak of $\Ks\sim10$\,mag
around $\Ks-[8.0]= 2$--4\,mag.  Oxygen-rich stars have slightly higher
luminosity in \Ks\, at a given colour, except for the two oxygen-rich
post-AGB/YSO candidates. In this diagram, oxygen-rich stars are well separated
from carbon-rich stars. LMC carbon-rich stars have HCN and C$_2$H$_2$
absorption bands at 7\,$\mu$m \citep{Aoki98, Jorgensen00, Matsuura06} in the
[8.0]-band (6.5--9.5\,$\mu$m) while oxygen-rich stars have only weak SiO
absorption and silicate excess in the [8.0]-band, until the silicate band
turns to absorption. These spectral differences separate the two chemical
types in this colour-magnitude diagram. This diagram also shows that high
mass-loss rate carbon-rich stars could have $\Ks$-band magnitudes fainter than
15\,mag, i.e., below the detection limit of the 2MASS $\Ks$-band.

\subsection{Other colour-magnitude diagrams}
 
\begin{figure}
\centering
\rotatebox{90}{ 
\begin{minipage} {6.5cm}
%\resizebox{\hsize}{!}{\includegraphics*[11,20][523,679]{mag58_8_lo.ps}}
\resizebox{\hsize}{!}{\includegraphics*[11,20][523,679]{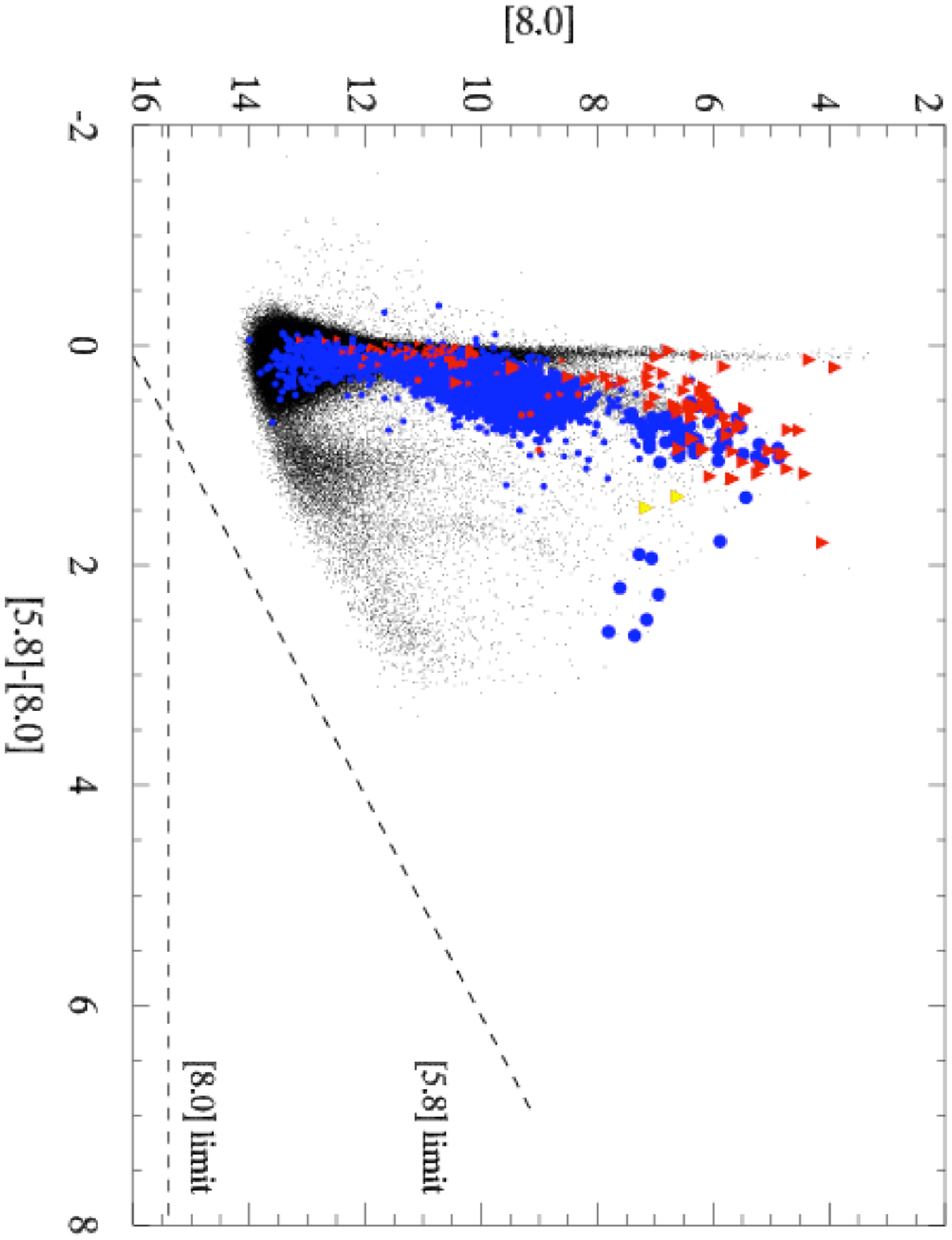}}
\end{minipage}}
\caption{
$[5.8]-[8.0]$ vs [8.0] colour-magnitude diagram.
Symbols are the same as in Fig.\ref{Fig-JK-K}.
Oxygen-rich and carbon-rich stars are found close together.
}
\label{Fig-58-8}
\end{figure}

\begin{figure}
\centering
\rotatebox{90}{ 
\begin{minipage} {6.5cm}
\resizebox{\hsize}{!}{\includegraphics*[11,20][523,679]{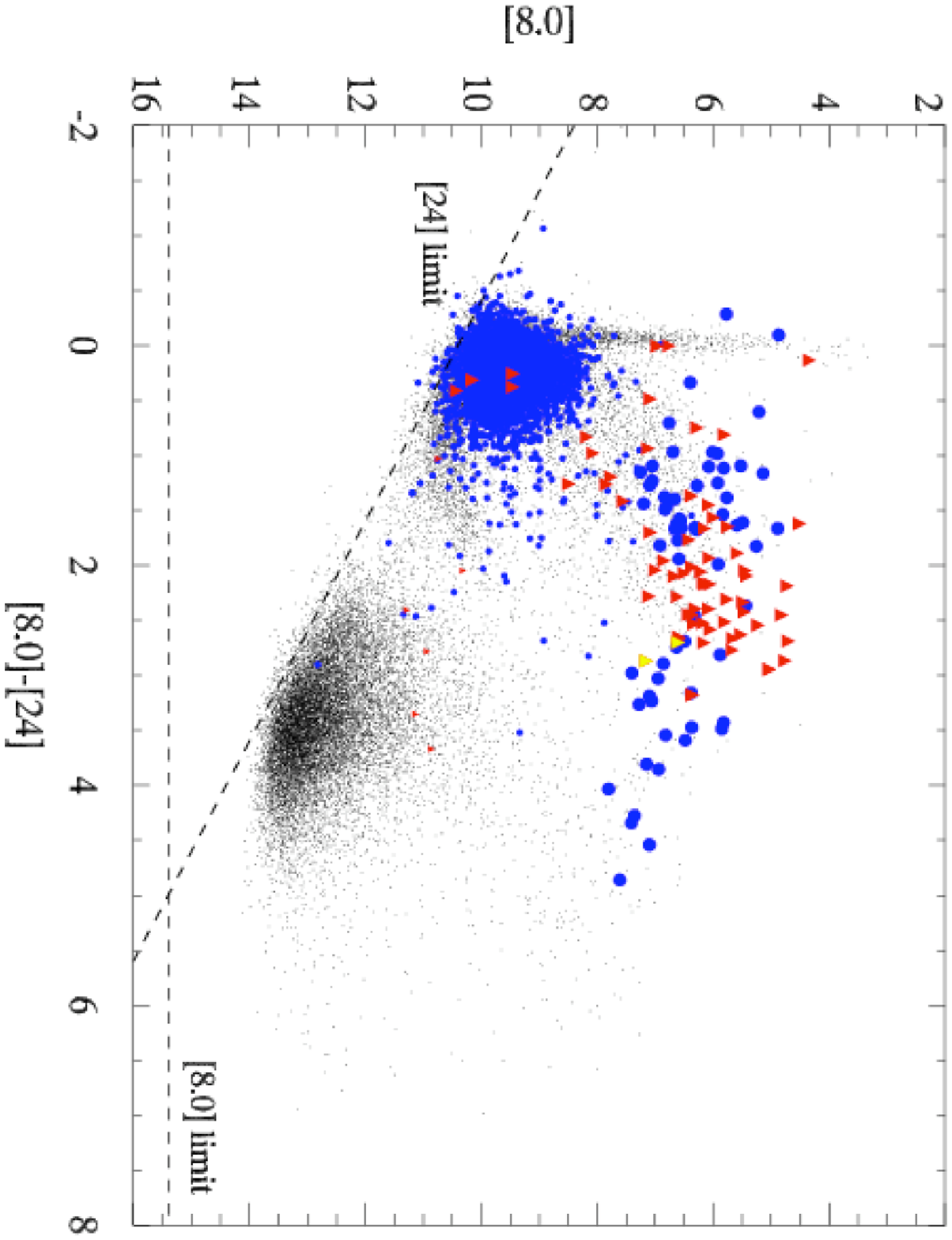}}
%\resizebox{\hsize}{!}{\includegraphics*[11,20][523,679]{mag824_8_lo.ps}}
\end{minipage}}
\caption{$[8.0]-[24]$ vs [8.0] colour-magnitude diagram.
Symbols are the same as Fig.\ref{Fig-JK-K}.
Oxygen-rich and carbon-rich stars are not separated.
}
\label{Fig-824-8}
\end{figure}

High mass-loss rate stars are likely to be detected at longer wavelength,
as is illustrated in the Spitzer SAGE colour-magnitude diagrams discussed
below. Figs.\,\ref{Fig-58-8} and \ref{Fig-824-8} show colour-magnitude
diagrams using the SAGE [5.8]-, [8.0]- and [24]-bands.
 In these two figures the sequence of IRS-classified carbon-rich stars
 does not end at the detection limits.
 Thus all of the red, i.e., high mass-loss rate carbon-rich, stars
 were detected in these bands.
However, there is no clear 
separation between oxygen-rich and carbon-rich stars so the 
diagrams are not useful for classification.
%_________________________________________________________________
%_________________________________________________________________
%_________________________________________________________________
%_________________________________________________________________

\subsection{Near-infrared survey with the IRSF}
\begin{figure}
\centering
\resizebox{\hsize}{!}{\includegraphics*[14, 7][477, 416]{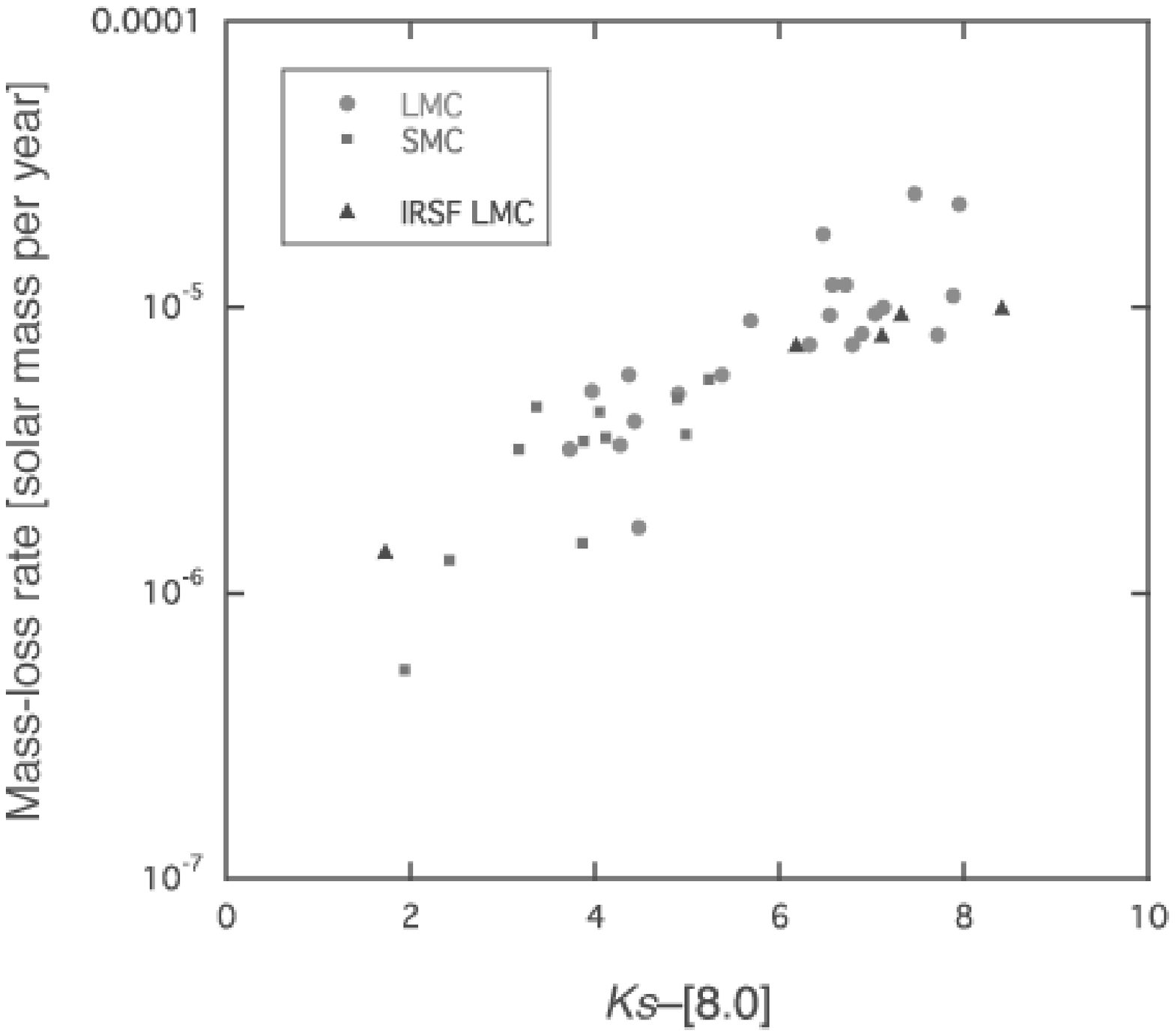}}
\caption{The \Ks$-$[8.0] using IRSF \Ks\, is plotted 
}
\label{Fig-K8-sirius}
\end{figure}

\citet{Kato07} published a $JHKs$ survey of the Magellanic Clouds taken with
the Infrared Survey Facility (IRSF) telescope at SAAO in South Africa. The
sensitivity of this survey, 16.6\,mag at 10\,$\sigma$ in the \Ks\,band for the
LMC, is significantly better than that of 2MASS (14.3\,mag), although the
conversion between IRSF and 2MASS filter system may not be straightforward for
AGB stars \citep{Kato07}. Four of the stars listed in \citet{Groenewegen07}
have been observed with IRSF and their mass-loss rate is plotted in
Fig.\,\ref{Fig-K8-sirius}. It appears that IRSF stars follow a similar colour
mass-loss rate trend. We will investigate this relation further after the
Spitzer spectroscopic survey, SAGE-Spec, is complete.

%%_________________________________________________________________
%\begin{figure}
%\centering
%\resizebox{\hsize}{!}{\includegraphics*[24, 138][486, 560]{sirius_2massK_JK.eps}}
%\caption{
%}
%\label{Fig-JK-Kdiff}
%\end{figure}
%%_________________________________________________________________
%_________________________________________________________________
%_________________________________________________________________

%_________________________________________________________________
\begin{figure}
\centering
\rotatebox{90}{ 
\begin{minipage} {6.5cm}
%\resizebox{\hsize}{!}{\includegraphics*[11,20][523,679]{mag324_3_lo.ps}}
\resizebox{\hsize}{!}{\includegraphics*[11,20][523,679]{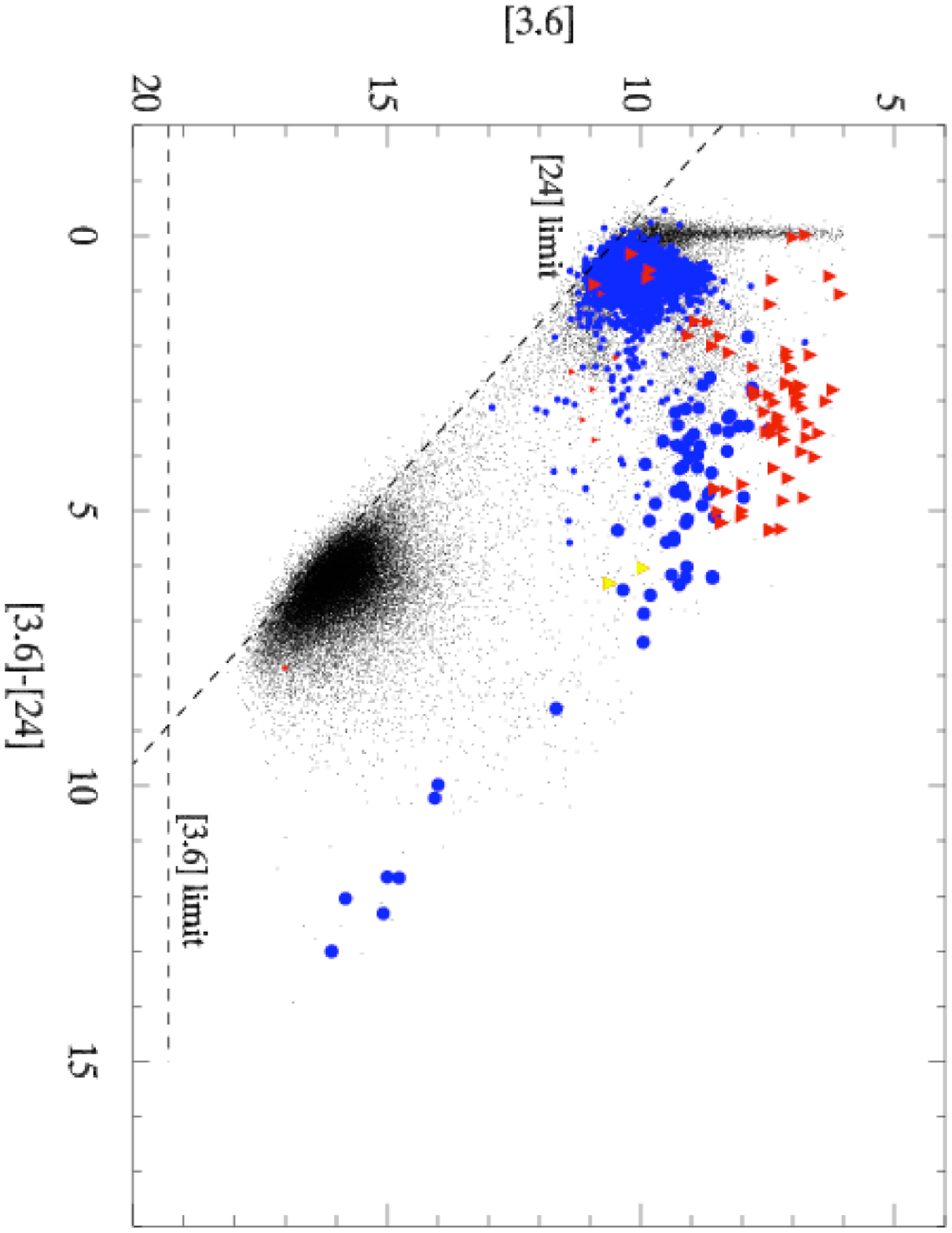}}
\end{minipage}}
\caption{[3.6]$-$[24] vs [24] colour-magnitude diagram.  
Symbols are the same as \ref{Fig-JK-K}.
}
\label{Fig-3-24}
\end{figure}
%_________________________________________________________________

\label{lastpage}

\end{document}